\documentclass[bibauthoryear]{aa}

\usepackage{amssymb}
\usepackage{soul}

\newcommand{\der}{\mathrm{d}}

\usepackage{graphicx}
\usepackage[varg]{txfonts}
\usepackage{amsmath}
\usepackage{natbib}
\usepackage{xcolor}
\usepackage[colorlinks=true,citecolor=blue,linkcolor=blue,urlcolor=blue,anchorcolor=blue]{hyperref}

\begin{document}

	\title{Cyclotron line formation in the radiative shock of an accreting magnetized neutron star}
	
	\author{Nick Loudas
		\inst{1,2,3}
		\and
		Nikolaos D. Kylafis\inst{1,2}
		\and
		Joachim Tr\"{u}mper\inst{4,5}
	}
	
	\institute{
        University of Crete, Department of Physics \& Institute of
		  Theoretical \& Computational Physics, 70013 Herakleio, Greece \\
		  \email{kylafis@physics.uoc.gr}
		\and
        Institute of Astrophysics, Foundation for Research and Technology-Hellas, 71110 Heraklion, Crete, Greece
        \and
        Department of Astrophysical Sciences, Peyton Hall, Princeton University, Princeton, NJ 08544, USA
		  \\
        \email{loudas@princeton.edu}
        \and
		  Max-Planck-Institut f\"{u}r extraterrestrische Physik, 
		  Postfach 1312, 85741 Garching, Germany
            \and
            University Observatory, Faculty of Physics, Ludwig-Maximilians Universit\"{a}t, Scheinerstr. 1, 81679 Munich, Germany\\
	    }
	
	\date{Received / accepted }
	
	\abstract
	{
 Magnetic neutron stars (NSs) often exhibit a cyclotron resonant scattering feature (CRSF) in their X-ray spectra. Accretion onto their magnetic poles is responsible for the emergence of X-rays, but the site of the CRSF formation is still an open puzzle. A promising candidate for high-luminosity sources has always been the radiative shock in the accretion column.  Nevertheless, no quantitative calculations of spectral formation at the radiative shock have been performed so far.
 }
	{
 It is well accepted that, in the accretion column of a high-luminosity accreting magnetic NS, a radiative shock is formed. Here we explore the scenario where the shock is the site of the cyclotron-line formation. We study spectral formation at the radiative shock and the emergent spectral shape across a wide range of the parameter space and determine which parameters hold an important role in shaping a prominent CRSF.    
 }
	{
 We developed a Monte Carlo code, based on the forced first collision numerical scheme, to conduct radiation transfer simulations at the radiative shock. The seed photons were due to bremsstrahlung and were emitted in the post-shock region. We properly treated bulk-motion Comptonization in the pre-shock region, thermal Comptonization in the post-shock region, and resonant Compton scattering in both regions. We adopted a fully relativistic scheme for the interaction between radiation and electrons, employing an appropriate polarization-averaged differential cross-section. As a result, we calculated the angle- and energy-dependent emergent X-ray spectrum from the radiative shock, focusing on both the CRSF and the X-ray continuum, under diverse conditions. The accretion column was characterized by cylindrical symmetry, and the radiative shock was treated as a mathematical discontinuity. 
 }
	{
 We find that a power law, hard X-ray continuum and a CRSF are naturally produced by the first-order Fermi energization as the photons criss-cross the shock. The depth and the width of the CRSF depend mainly on the transverse optical depth and the post-shock temperature. We show that the cyclotron-line energy centroid is shifted by $\sim(20-30)\%$ to lower energies compared to the classical cyclotron energy, due to the Doppler boosting between the shock reference frame and the bulk-motion frame. We demonstrate that a "bump" feature arises in the right wing of the CRSF due to the upscattering of photons by the accreting plasma and extends to higher energies for larger optical depths and post-shock temperatures. 
 }
	{
 We conclude that resonant Compton scattering of photons by electrons in a radiative shock is efficient in producing a power-law X-ray continuum with a high-energy cutoff accompanied by a prominent CRSF. The implications of the Doppler effect on the centroid of the emergent absorption feature must be considered, if an accurate determination of the magnetic field strength is desired. 
 }
	

	\authorrunning{N. Loudas et al.}
	\titlerunning{Cyclotron line formation in a radiative shock} 
	
	\maketitle
	
	
\section{Introduction} \label{sec1}

	Accretion-powered X-ray Pulsars (XRPs) are strongly magnetized neutron stars (NSs) that accrete matter from a donor companion star. Due to the NS's enormous magnetic field ($B \sim 10^{12}\,\mathrm{G}$), the inner region of the accretion disk is disrupted at a distance, which is called the Alfven radius, and the plasma gets attached to the field lines. From there, the accreting matter is channeled onto the magnetic poles of the NS, where most of the gravitational energy is released, thereby resulting in the emergence of X-rays (\citealt{Basko1976}). For a recent review, see \cite{Mushtukov2022}.

    A quantum-mechanical phenomenon, called magnetic resonant scattering, gives the distinct opportunity to measure precisely the NS's magnetic field strength on its surface. The motion of a charged particle (e.g., electron) in the presence of a strong magnetic field is quantized across its axis (\citealt{Landau1930}; see also \citealt{landau1965}), leading to discrete energy states that are commonly-known as Landau levels. Hence, absorption/emission features are expected to appear in the spectra of highly magnetized NSs. These features are the so-called cyclotron lines, also known as cyclotron resonant scattering features (CRSFs). The difference in energy between two successive levels encodes information about the local magnetic field strength existing in the line-forming region, as it is directly proportional to its strength. The detection of such a line in the spectrum of a NS offers a direct and robust way of measuring the magnetic-field strength in NSs (\citealt{Gnedin1974, Basko1975}). On top of that, cyclotron lines are thought to be a solid diagnostic tool to probe the physics of accretion onto the magnetic poles of highly magnetized NSs (\citealt{Staubert2019}).
    
    The first ever discovery of a cyclotron line in NSs occurred forty-six years ago in the X-ray spectrum of the accreting binary pulsar Hercules X-1 (\citealt{Truemper1977,Truemper1978}), revealing this new powerful phenomenon to the X-ray community. Since then, more than 35 electron cyclotron lines have been detected in accreting magnetic NS X-ray spectra, sometimes with their harmonics (for a review see, \citealt{Staubert2019}) and even more rarely with anharmonics (see e.g., \citealt{Fuerst2018,Sharma2022,Yang2023}), covering an energy range from $\sim 10~\mathrm{keV}$ to $\sim 100~\mathrm{keV}$. 
    
    A large amount of observational data has been collected for most of these sources. Yet, the region where the CRSFs are formed, as well as the mechanism that produces them remain an open fundamental puzzle, preventing us from inferring the NS's magnetic field strength accurately, as it varies noticeably on the NS surface and along the accretion column, as well. 
	
	In the standard picture, the formation of cyclotron lines in NSs is thought to take place in the lower part of the accretion column through resonant scattering (\citealt{Gnedin1974, Basko1975}). The accretion flow heats the magnetic polar cap resulting in a hot-spot that emits soft X-ray photons. If the accretion rate is high enough, above a critical value, then the radiation pressure exerted on the in-falling plasma is efficient in decelerating the material well before it hits the surface of the NS, thereby giving rise to a radiative shock (for a detailed discussion see \citealt{Mushtukov2015a}). While passing through the shock, the in-falling material slows down transferring most of its gravitational energy to photons via collisions of each electron with a multitude of photons and eventually settling down to the post-shock region, having relatively small flow velocity and mainly thermal one. \cite{Kylafis2014} demonstrated that this mechanism is capable of producing X-ray power-law spectra through a combination of bulk and thermal Comptonization of bremsstrahlung seed photons in the shock. \cite{Basko1976} suggested that a fraction of the radiation, that has energy comparable to the cyclotron energy, will undergo resonant scattering while criss-crossing the shock, eventually producing a CRSF in the emergent spectrum. Yet, no calculation has been performed so far for the formation of an X-ray power-law spectrum along with a cyclotron line in a radiative shock.
	
	Over the last decades, several extensions to this scenario and other models have been proposed to address the problem of CRSF formation in NSs. They can be divided into two different classes: the first one postulates that the absorption feature emerges in the accretion column (e.g., \citealt{Becker2007, Becker2012, Nishimura2014, Musthukov2015b,Becker2022}), while the latter considers the reprocessing of X-rays emitted by the accretion column on the NS surface (see, e.g., \citealt{Poutanen2013}). Nevertheless, none of them has been confirmed to be quantitatively correct yet. 
	
	An observation that could be instrumental in constraining the currently viable models is the variation of the cyclotron line energy with the emergent X-ray luminosity. Several CRSFs, reported in XRP spectra, exhibit a correlation between the line energy $E_{c}$ and the emergent X-ray luminosity $L_{X}$, indicating that the configuration of the line-forming region may depend on the accretion rate. It has been observed that XRPs of relatively low luminosity show a positive $E_{c} - L_X$ correlation (e.g., Her X-1: \citealt{Staubert2007}; GX 304-1: \citealt{Klochkov2012}; Vela X-1:  \citealt{Fuerst2014}). In contrast, sources of high luminosity often exhibit a negative correlation (e.g., V0332+53: \citealt{Tsygankov2006}). This trend suggests that whether the correlation is positive or negative is dictated by a critical value of the X-ray luminosity ($L_{cr} \sim 10^{37}\,\mathrm{erg~s^{-1}}$) above which the correlation becomes negative. This phenomenon has already been observed in two XRPs (1A 0535+262: \citealt{Kong2021,Shui2024} and V 0332+53: \citealt{Doroshenko2017,Vybornov2018}), but see also the GRO J1008-57 \citep{Chen2021}. 
	
	\cite{Basko1976} were the first to show that there exists such a critical value for the accretion luminosity above which the formation of a radiative shock takes place, stating that the wide range of observed luminosities in accreting XRPs is likely associated with different accretion regimes. This threshold naturally splits the accretion regimes into two distinct types: 1) accreting NSs of sub-critical accretion luminosity with no shock at all (but see \citealt{Bykov2004}), and 2) XRPs of super-critical luminosity that can support a radiation-dominated shock in the accretion column.
	 
	Sub-critically accreting NSs have been studied extensively and a solid model was proposed recently by \cite{Musthukov2015b}. According to this model, the deceleration of the in-falling material occurs on the NS surface, at a height comparable to the size of the hot-spot, via scattering of electrons by the hot-spot's radiation. This scenario invokes the Doppler effect to explain the positive correlation between the emergent X-ray luminosity $L_X$ and the cyclotron line centroid $E_c$, due to the variation of the velocity profile with radiation pressure. It has been successful in describing the data of the XRP GX 304-1. Alternative scenarios include: 1) braking of the plasma solely by Coulomb collisions near the NS atmosphere proposed by \cite{Staubert2007}, and 2) emergence of a strong collisionless shock (CS) above the polar cap that dissipates the ion's kinetic energy (\citealt{Bisnovatyi1970, Shapiro1975, Langer1982, Bykov2004}, for an implementation to observations see \citealt{Rothschild2017, Vybornov2017}). In the Coulomb collisions (collisionless shock) model, the larger the luminosity, the smaller the characteristic ion's stopping length (shock height) and therefore the greater the local magnetic field strength (i.e., cyclotron energy) in the line-forming region. This naturally yields a positive $E_c - L_X$ correlation.
	
	On the other hand, accreting NSs that exhibit a super-critical luminosity are generally described by the model of \cite{Basko1976}. In this scenario, the height of the shock scales approximately linearly with the X-ray luminosity, while the magnetic field strength drops with the increase of the shock's height. Thus, it predicts an anti-correlation between the line energy and the X-ray luminosity, qualitatively in line with the observations of super-critical XRPs. However, it was  pointed out by \cite{Poutanen2013} that in a dipole field, the predicted rate of change of $E_{c}$ with $L_X$ is much larger than the observed one. As a remedy, the authors put forward the idea that the observed cyclotron line is not produced in the shock, but from the reflection of radiation produced in the shock in the area surrounding the polar cap. 
	
	In the recent work of \cite{Kylafis2021}, we investigated the cyclotron-line formation scenario proposed by \cite{Poutanen2013}. The reprocessing of photons on the NS surface has, in principle, the ability to produce a negative correlation between cyclotron-line centroid and X-ray luminosity.  As the luminosity increases, the shock height increases also and thus more and more photons are scattered at lower and lower magnetic latitudes, where the magnetic field is lower.  The magnetic field at the equator is half that of the pole. We employed a Monte Carlo (MC) code to derive the reflected spectrum and we demonstrated that this mechanism cannot produce prominent CRSFs in the emergent X-ray spectrum similar to the ones observed in NS spectra. We also showed that the reflection model cannot explain the negative correlation between the cyclotron energy and X-ray luminosity reported in the emergent spectrum of V0332+53 source (\citealt{Tsygankov2010}). The reason is that scattering preserves the number of photons.  Thus, photons that are missing in an absorption feature will appear as an emission feature, a "bump", at slightly lower energy.  Adding the spectra from all illuminated magnetic latitudes washes out the expected correlation.

	In this work, we aim to revisit the model of \cite{Basko1976} and explore the possibility of cyclotron lines being formed in the radiative shock in the accretion column of a NS by means of a MC simulation. We consider an accreting magnetic NS being in the super-critical regime with a radiative shock in a cylindrical accretion column. Our input photons are bremsstrahlung ones emitted below the shock. We let the photons criss-cross the shock as many times as they want before they escape sideways. We derive the emergent spectrum from the accretion shock taking into account both thermal- (post-shock) and bulk-Comptonization (pre-shock), as well as the resonant Compton scattering of photons by electrons, implementing the prescription developed in \cite{Loudas2021}, thereby allowing for the emergence of a CRSF in the total spectrum. The importance of the Doppler effect between the shock's reference frame and the bulk-motion's frame in the CRSF's energy centroid is also quantified. This model is an extension of the one proposed by \cite{Kylafis2014} to explain the hard X-ray power-law tails observed in the spectra of anomalous X-ray pulsars (AXPs) and soft gamma-ray repeaters (SGRs).
	
	The layout of this paper is as follows. In Sect. \ref{sec2}, we offer a qualitative presentation of our model, giving special emphasis on the physical processes that occur in the radiative shock and offering useful scaling relations among the parameters considered in this study. In Sect. \ref{sec3}, we describe the Monte Carlo code that is implemented in this study. In Sect. \ref{sec4}, we exhibit our results, providing a qualitative comparison with observations. In Sect. \ref{sec5}, we comment on the results and draw our conclusions.

\section{The model} \label{sec2}
	 
To demonstrate our underlying ideas, we use a relatively simple model for describing accreting NSs of super-critical luminosity. Let us consider a highly magnetized NS that is accreting matter
at an accretion rate $\dot M$. The strong dipole magnetic field channels the plasma to the magnetic poles of the NS. For simplicity, we assume the accretion column to be a  cylinder of cross-sectional area $\pi a^2$ (though accretion physics dictates that the accretion column is not cylindrical, but bow-like around the magnetic pole, see e.g., \citealt{Basko1976,Romanova2004}) and a radiative shock at height $H$ above the magnetic pole.  Above the shock, matter falls freely at a constant speed (pre-shock region), while below the shock (post-shock region) it has negligible flow velocity, electron temperature $T_e$, and density seven times the pre-shock one. The free-falling plasma is decelerated by the radiation field while passing through the shock, which is treated here as a mathematical discontinuity. To fit real data, more detailed models are required.

Bremsstrahlung photon beams from below the shock are injected isotropically upward and each beam is followed by our Monte Carlo code (see Sect. \ref{sec3}) until its intensity has dropped below $0.1\%$ of the injected value. The crucial parameters in this calculation are the free-fall velocity, the accretion luminosity (i.e., accretion rate), the magnetic field strength at the shock, and the transverse Thomson optical depth, since the longitudinal optical depth in both the pre-shock and the post-shock regions is infinite by construction in our model.

\subsection{Qualitative considerations} \label{sec2.1}

\subsubsection{Accretional energy}
\label{sec2.1.1}
Let's assume for demonstration purposes that the accreting plasma is composed only of protons and electrons. Due to charge and current neutrality, each free-falling proton is accompanied by an electron that travels at the same speed. Thus, all the accretional energy resides in the protons. Ignoring small thermal motion, we can treat all proton-electron pairs as if they are accreting at speed 
\begin{equation}
	v_{ff} = \sqrt{2GM/r},
	\label{6.0}
\end{equation}
where $G$ is the gravitational constant, $M$ is the NS's mass, $r$ is the distance from its center, i.e., $r= R + H$, where $R$ is the NS's radius. In a typical NS, the infalling matter can reach mildly relativistic speed, e.g., $v_{ff} \sim 0.5 ~c$. Each pair at $r>R+H$ has energy 
\begin{equation}
E_{ff} = E_p + E_e \approx GMm_p / r.
\label{6.1}
\end{equation}
For a typical NS, this energy is of the order of $10^2$ MeV, well above the X-ray band.

\subsubsection{Deceleration mechanism \& shock parameters}
\label{2.1.2}

At first sight, the accretion flow's deceleration seems impossible to occur in the radiative shock, as the material below the shock is thermal and emits Bremsstrahlung radiation, which, in photon number, dominates in the soft X-ray band. For simplicity, let's assume that initially, each proton-electron pair has energy $E_{ff}$ of the order of $100$ MeV, while the X-ray photons have energies $E_\gamma$ that lie in the range $(1-100)$ keV. After a head-on collision of a photon with a relativistic electron, the photon-energy gain is at most $\Delta E_\gamma \approx \text{(a few)} \times E_\gamma$, which is way smaller than the proton-electron pair's accretional energy (see Eq. \ref{6.1}). Hence, the pair loses only a tiny fraction of its accretional energy in a scattering. As we demonstrate below, though, the electron's mean free path is orders of magnitude smaller than the photon's mean free path. Thus, each electron gets scattered by many photons while passing through the shock. This mechanism makes it possible for the accretion flow to transfer most of its energy to photons and slow down. The phenomenon is similar to the slowing down of a diver by the water molecules in the top layers of water in a pool.

We envision the deceleration process as follows: 
In-falling electrons undergo multiple scatterings with the X-ray photons that try to escape the accretion column upward, and hence they gradually lose energy, while radiation escapes sideways. But, charge neutrality implies that after each scattering event between a photon and an electron, a `companion' proton supplies a small fraction of its energy to the electron in order to counterbalance its loss due to scattering. As a result, the electron acts as a mediator to transfer the proton's energy to photons while traversing the shock. This process takes place in the accretion shock and it is terminated when the protons and the electrons acquire the same temperature behind the shock. Afterwards, the plasma sinks down in the post-shock region, having negligible flow velocity, but only thermal one, and losing energy via bremsstrahlung. A fraction of the bremsstrahlung photons goes upward, creates the shock, and in the process the upscattered photons that escape constitute the largest fraction of the observed luminosity, called accretion luminosity, which equals 
\begin{equation}
L_{acc} = G\dfrac{M\dot{M}}{r} =  \dfrac{1}{2} \dot M \beta_{ff}^2 c^2,
\label{6.2}
\end{equation} 
where $\beta_{ff}$ is the free-fall plasma velocity $v_{ff}$ normalized to the speed of light $c$. 

The photons' energy density $u_\gamma$ in the shock region is estimated by
\begin{equation}
	u_\gamma = \dfrac{L_{acc}}{\pi a^2 c}.
	\label{6.3}
\end{equation}
It is convenient to express $u_\gamma$ in terms of the mass accretion rate, as it can then be estimated directly by invoking the continuity equation. Substituting Eq. \eqref{6.2} into Eq. \eqref{6.3}, we get
\begin{equation}
	u_\gamma = \dfrac{\dot M}{\pi a^2}\dfrac{v_{ff}^2}{2c}.
	\label{6.4}
\end{equation}
Conservation of mass yields the following approximate formula for the plasma density in the shock region
\begin{equation}
	\rho = \dfrac{\dot M}{\pi a^2 v_{ff}}, 
	\label{6.5}
\end{equation}
which, after taking into account charge neutrality (implying $n_e = n_p$), as well as that $m_{e}\ll m_{p}$, becomes
\begin{equation}
	n_e = \dfrac{\dot M/m_p}{\pi a^2 v_{ff}}.
	\label{6.6}
\end{equation}
Using Eq. \eqref{6.4} along with Eq. \eqref{6.6}, we find
\begin{equation}
	\dfrac{u_\gamma}{n_e} = \beta_{ff}\dfrac{1}{2}m_p v^2_{ff}.
	\label{6.7}
\end{equation}
Next, writing $u_\gamma$ as the photons' number density $n_\gamma$ times a characteristic photon's energy $E_\gamma$, the ratio $n_\gamma/n_e$ reads
\begin{equation}
\dfrac{n_\gamma}{n_e} = \beta_{ff} \dfrac{m_p v^2_{ff}}{2E_\gamma} =\beta_{ff} \dfrac{E_{ff}}{E_\gamma} , \label{6.9b}
\end{equation}
where $E_{ff}$ is given by Eq. \eqref{6.1}.

The characteristic mean free path for propagation of electrons in the `bath' of outgoing photons is
\begin{equation}
	\bar \lambda_e = \dfrac{1}{n_\gamma \sigma_\tau},
	\label{6.9}
\end{equation}
while the corresponding photon's mean free path is given by
\begin{equation}
	\bar\lambda_{\gamma} = \dfrac{1}{n_e \sigma_\tau},
	\label{6.8}
\end{equation}
where, for simplicity in these back-of-the-envelope calculations, we employ the classical Thomson cross-section $\sigma_\tau$ for the $e-\gamma$ interaction. We note that in our Monte Carlo calculation, we make use of the accurate magnetic resonant Compton cross-section, which accounts for the photon's direction of propagation, energy, and polarization and depends on the magnetic field (see Sect. \ref{sec3.4}).

Combining Eqs. \eqref{6.9b}, \eqref{6.9}, \& \eqref{6.8}, we find the ratio $\bar\lambda_{\gamma}/\bar\lambda_e$ to be
\begin{equation}
    \dfrac{\bar\lambda_{\gamma}}{\bar\lambda_e} = \beta_{ff}\dfrac{E_{ff}}{E_\gamma} \lesssim E_{ff} / E_\gamma \sim {\cal O}(10^3 - 10^4) \label{6.8b}
\end{equation}
where we used that $\beta_{ff} \lesssim 1$. Eq. \eqref{6.8b} indicates that the electron's mean free path is significantly smaller than the photon's one, as the number density of photons is way greater than the electrons' one (see Eq. \ref{6.9b}). Given that a characteristic length scale for the shock's width is the mean free path of the photons (see Eq. \ref{6.12}), this result implies that each electron gets scattered by many photons in the shock. Since the ratio of the photon mean free path to the electron's one is approximately inversely proportional to their corresponding energies ratio, the effective deceleration of the free-falling plasma in the shock arises naturally, in line with the statement we made above.

By construction in our model, the accretion column is effectively infinite along the magnetic field's axis; thus, the longitudinal optical depth for the photon's propagation is infinite, implying that all photons escape sideways. So, an important quantity for our calculations is the transverse Thomson optical depth $\tau_\perp$, which is given by 
\begin{equation}
	\tau_\perp = n_e \sigma_\tau a = a / \bar\lambda_{\gamma}.
	\label{6.10}
\end{equation}    
Substituting Eq. \eqref{6.6} into \eqref{6.10}, we obtain a relation between the transverse optical depth and the mass accretion rate, namely
\begin{equation}
	\tau_\perp = \dfrac{\dot M}{m_p\pi a v_{ff}} \sigma_\tau.
	\label{6.11}
\end{equation}

Finally, we can infer a characteristic value for the longitudinal width of the radiative shock as follows: the shock is roughly defined as the region in which the in-falling plasma loses most of its energy, transferring it to photons through multiple scatterings. As long as $E_{ff}\gg E_\gamma$, after each scattering event the proton-electron pair's energy is reduced by an amount $ \Delta E \sim 3 E_\gamma$ (if $\beta_{ff} = 0.5$), on average. Thus, $N \sim E_{ff}/E_\gamma$ scattering events are required for an electron-proton pair to be effectively decelerated, supplying its energy to X-ray photons. We demonstrated earlier on that the ratio $E_{ff}/E_\gamma$ is roughly equal to $n_\gamma / n_e$ (see Eq. \ref{6.9b}). Hence, if each photon has undergone one scattering off an electron, on average, then each electron will have been scattered $\sim n_\gamma / n_e$ times, on average, converting in this way the accretional energy into X-ray radiation for the reasons mentioned above. This simple consideration leads us to define the characteristic shock's width, $\zeta$, as the longitudinal mean free path of the photons $\bar\lambda_{\gamma}$ given in \eqref{6.8}, since it represents the statistically average distance that photons travel between two sequential scattering events. Therefore,
\begin{equation}
	\zeta \simeq \bar\lambda_{\gamma} = \dfrac{1}{n_e \sigma_\tau}.
		\label{6.12}
\end{equation}
For the transverse size of the shock across the column, we adopt the work of \cite{Kylafis2014}. They offer the following simple formula for the radius of the accretion column at the shock 
\begin{equation}
	a \approx 270 \left(\dfrac{\dot M}{10^{16}\, \mathrm{g \, s^{-1}}}\right)^{1/5}
	\left(\dfrac{B_0}{10^{12}\, \mathrm{G}}\right)^{-1/4}\,\mathrm{m},
	\label{6.13}
\end{equation}
where $B_0$ is the magnetic field strength on the NS surface.

\subsection{Convenient expressions} \label{sec2.2}

In Sect. \ref{sec2.1}, we derived simple formulae for every parameter that is considered important in studying the radiation that emerges from the radiative shock in the accretion column. It is helpful to derive characteristic values for these parameters.  For a typical accreting magnetized NS, we have $M= 1.4 \,\mathrm{M_\odot}, \, r = 10 \,\mathrm{km} \text{ (assuming }H\ll r\text{) }, ~ B_0 = 10^{12} \, \mathrm{G}, \text{and } \dot M = 5 \times 10^{16} \,\mathrm{g \, s^{-1}}$. Thus, Eqs. \eqref{6.0}, \eqref{6.2}, \eqref{6.11}, \& \eqref{6.12} are for convenience written as
\begin{equation}
	\beta_{ff} \simeq 0.65 \left(\dfrac{M}{1.4~ \mathrm{M_\odot}}\right)^{1/2}\left(\dfrac{r}{10 \,\mathrm{km}}\right)^{-1/2},\label{6.14}
\end{equation}
\begin{equation}
	L_{acc} \simeq 2\times10^{36} \left(\dfrac{\dot M}{ 10^{16}~\mathrm{g~s^{-1}}}\right)\left(\dfrac{M}{1.4~ \mathrm{M_\odot}}\right)
	\left(\dfrac{r}{10 \,\mathrm{km}}\right)^{-1} \, \mathrm{erg ~ s^{-1}},
	\label{6.15}
\end{equation}
\begin{equation}
	\tau_\perp \simeq 2.5 \left(\dfrac{\dot M}{ 10^{16}~\mathrm{g~s^{-1}}}\right)^{4/5}
	\left(\dfrac{M}{1.4~ \mathrm{M_\odot}}\right)^{-1/2}\left(\dfrac{r}{10 \,\mathrm{km}}\right)^{1/2}
	\left(\dfrac{B_0}{10^{12}\, \mathrm{G}}\right)^{1/4},
	\label{6.16}
\end{equation}
\begin{multline}
	\zeta \simeq 107 
	\left(\dfrac{\dot M}{ 10^{16}~\mathrm{g~s^{-1}}}\right)^{-3/5}
	\left(\dfrac{M}{1.4~ \mathrm{M_\odot}}\right)^{1/2}\\
    \times \left(\dfrac{r}{10 \,\mathrm{km}}\right)^{-1/2}
	\left(\dfrac{B_0}{10^{12}\, \mathrm{G}}\right)^{-1/2}~\mathrm{m}. \label{6.17}
\end{multline}
Note that for high-luminosity sources ($\dot M \sim 10^{17}~ \mathrm{g~ s^{-1}}$), we are allowed to consider $\zeta \ll H$.


\section{The Monte Carlo code} \label{sec3}

	The Monte Carlo (MC) code that is employed in this study is an extension of the one used in \cite{Kylafis2014}, with the use of resonant cross sections (see Sect. \ref{sec3.4} below). The inner workings of the code are based on the works of \cite{Cashwell1959} and \cite{Pozdnyakov1983} and the radiative transfer method is the so-called \textit{forced first collision}, as it allows us to treat properly the abrupt change in the mean free path from the pre-shock region to the post-shock one and results in high-accuracy even for optically thin media (i.e., $\tau_\perp \ll 1$). For a recent review on MC radiative transfer schemes see \cite{Noebauer2019}.
	
	\subsection{Accretion column: geometry \& plasma}
    \label{sec3.1}	
	We assume a radiative shock with infinitesimal width at height $H$, above the NS surface, in a cylindrical accretion column of radius $a$ (see Eq. \ref{6.13}). The shock is of the Rankine-Hugoniot form. Above the shock (pre-shock region; $z>H$) matter falls freely at a constant speed (see Eq. \ref{6.14}), while in the post-shock region, below the shock ($z<H$), the density of the matter is increased by a factor of 7 (appropriate for a radiative shock), hot, and has negligible flow velocity. It should be noted that the downstream sinking velocity would be $v_{d} = v_{ff}/7 \lesssim 0.1 ~\mathrm{c}$. However, the rms in the velocity due to thermal motion is $v_{rms} = \sqrt{kT_e/m_ec^2} ~ \mathrm{c} \approx 0.1 (kT_e / 5~ \mathrm{keV})^{1/2} ~\mathrm{c}$. Thus, for temperatures considered in this study, the rms velocity is greater than the sinking flow velocity in the post-shock region and thus electrons motion is mainly thermal. Even if downstream electrons were much colder, the effect of this non-relativistic flow would be minimal, as it corresponds to a Lorentz factor of $\gamma \approx 1.005$. In essence, the fractional shift in the cyclotron energy due to Doppler would be $\sim 0.5 \%$, which is definitely negligible for the scope of this study. Therefore, we can safely ignore the sinking motion in the post-shock region. 
 
    The temperature of electrons in the post-shock region is not well determined from theory, though it is expected to be proportional to the kinetic accretional energy of electrons (see \citealt{Kylafis2014}). Thus, we consider various values of $kT_e$ in this study (see Sect. \ref{sec4.3}). The energy (or better the Lorentz factor) distribution of the thermal electrons in the post-shock region is taken to be the relativistic Maxwell-J\"{u}ttner one, namely
	\begin{equation}
		f_e(\gamma,T_e)~ \mathrm{d}\gamma = \dfrac{\gamma \sqrt{\gamma^2 - 1}}{\vartheta K_2(1/\vartheta)} \exp\left(-\gamma/\vartheta\right)~\mathrm{d} \gamma ,
		\label{3.1}
	\end{equation}
where $\gamma = 1 / \sqrt{1 - (v/c)^2}$ is the Lorentz factor, $v$ is the electron velocity, $K_2(x)$ is the modified Bessel function of the second kind and second order, and $\vartheta = kT_e / m_e c^2$.

	\subsection{Seed photons}
	\label{sec3.2}
	Seed photon beams from below the shock are injected upward, and each of them is followed by the Monte Carlo code until its intensity drops by at least a factor of a thousand, that is, $99.9\%$ of the injected beam escapes. The modeling of their injection is as follows:  photons are emitted isotropically upwards from a disk at $z=H$, that is we do not perform radiative transfer on the isotropically
	emitted seed photons in the post-shock region,
	but instead assume an upward flux of such photons. 
	
	The input photon energy distribution is considered to be non-magnetic Bremsstrahlung \citep{Becker2007}, namely
	\begin{equation}
		E\dfrac{\mathrm{d}N}{\mathrm{d}E} \propto G(E/2kT_e) \exp(-E/kT_e),\label{3.2}
	\end{equation}
	where 
	\begin{equation}
		G(x) = e^x K_0(x), \label{3.3}
	\end{equation}
	is the Gaunt factor in the Born approximation for the cross-section \citep{Greene1959}, $E$ is the seed photon energy, and $K_0(x)$ is the modified Bessel function of the second kind, zeroth order.

	\subsection{Mean free path} \label{sec3.3}
	
	A proper treatment of the mean free path is required in order for accurate radiative transfer calculations to be performed. Since we have a two zone model, the mean free path does not only depend on the direction of propagation and on the incident photon energy, but it varies significantly from the pre-shock region to the post-shock one.
	
	The mean free path for every position in the accretion column and in any direction is given by (\citealt{Pozdnyakov1983, Schwarm2017})
	\begin{equation}
		\left< \lambda(E,\theta) \right>_{f_e} = \dfrac{1}{n_e \left< \sigma(E,\theta) \right>_{f_e}}, \label{3.4}
	\end{equation}
	where $E,~\theta$ are the incident photon energy and angle with respect to the magnetic field axis, respectively, as measured in the lab (shock) reference frame, $n_e$ is the local number density of electrons, and $\left<\sigma(E,\theta)\right>_{f_e}$ is the averaged (over electron speeds) magnetic resonant Compton cross-section, given by
	\begin{equation}
	\left<\sigma(E,\theta)\right>_{f_e} = \dfrac{\int_{-1}^{1}f_e(\beta) \,(1 - \beta \cos\theta)\,\sigma(E_{rf},\theta_{rf})\,\mathrm{d}\beta}{\int_{-1}^1 f_e(\beta)\,\mathrm{d}\beta},
	\label{3.5}
	\end{equation}
	where $f_e(\beta)$ refers to the electron's velocity distribution function (see Eq. \ref{3.1}), while the subscript `$rf$' stands for quantities that are measured in the electron's rest frame and can be easily obtained through Lorentz transformations from the lab frame to the electron's one, and $\sigma(E,\theta)$ is the scattering cross-section (see Sect. \ref{sec3.4}). Following \cite{Harding1991} (see also \citealt{Schwarm2017}), we average the cross-section over only
	the longitudinal electron temperature, that is we presume the motion of electrons to be effectively along the magnetic field axis; thus the integration in \eqref{3.5} is over the electron velocity distribution along the field axis.
	
	In the pre-shock region, the electron velocity distribution function is essentially a Dirac $\delta-$function, i.e., $f_{e}(\beta) = \delta(\beta - \beta_{ff})$ and therefore the calculation of the mean free path (Eq. \ref{3.4}) is trivial. In the post-shock region, the electron velocity function is related to \eqref{3.1} and thus the integral is carried out numerically (see Appendix \ref{App.A}).   
	
	\subsection{Resonant Compton scattering prescription}
	\label{sec3.4}
	
	The radiative transfer calculation requires accurate expressions for the description of the radiation-matter interaction. The complete, quantum electrodynamical, Compton cross-sections are thoroughly discussed in the literature \citep[e.g.,][]{Harding1991,Sina1996,Nobili2008,Gonthier2014,Mushtukov2016}, but they are rather cumbersome and their implementation to MC codes turns out to be a computationally expensive task. In a previous paper \citep{Loudas2021}, we offered a simple, but accurate prescription for resonant Compton scattering calculations along with approximate expressions for the polarization-dependent differential cross-sections, which are valid for photons with energy $E\ll m_ec^2$ and for cyclotron line energies $E_{c} \ll m_e c^2$, that is for magnetic field strengths $B\ll B_{cr} \equiv m_e^2c^2/e\hbar = 4.41\times 10^{13}~\mathrm{G}$. We employ this prescription in our MC code, as most of the cyclotron lines that have been observed are at cyclotron energies $\ll m_ec^2$ \citep{Staubert2019},
	and therefore our work is applicable to most of the observed accreting NSs that exhibit a CRSF in their spectra. For CRSFs at energies $\sim m_ec^2$, the implementation of the complete cross-section expressions is inevitable.
	
	The approximate polarization-dependent resonant Compton cross-sections read \citep{Loudas2021}
	
	\begin{equation}
	{ {d \sigma_{11}} \over {d\Omega^\prime} } \approx
	{ {3 \pi r_0 c} \over 8} 
	\left(
	\cos^2\theta_{rf} \cos^2 \theta_{rf}^\prime ~ L_- + 
	{1 \over 2} {B \over {B_{cr}} } ~ L_+
	\right), \label{3.6}
	\end{equation}
	\begin{equation}
	{ {d \sigma_{12}} \over {d\Omega^\prime} } \approx
	{ {3 \pi r_0 c} \over 8} 
	\left(
	\cos^2\theta_{rf} ~ L_- + 
	{1 \over 2} {B \over {B_{cr}} } \cos^2\theta_{rf}^\prime ~ L_+
	\right), \label{3.7}
	\end{equation}
	\begin{equation}
	{ {d \sigma_{21}} \over {d\Omega^\prime} } \approx
	{ {3 \pi r_0 c} \over 8} 
	\left(
	\cos^2 \theta_{rf}^\prime ~ L_- +
	{1 \over 2} {B \over {B_{cr}} } \cos^2\theta_{rf} ~ L_+
	\right), \label{3.8}
	\end{equation}
	\begin{equation}
	{ {d \sigma_{22}} \over {d\Omega^\prime} } \approx
	{ {3 \pi r_0 c} \over 8} 
	\left(
	L_- + {1 \over 2} {B \over {B_{cr}} } \cos^2\theta_{rf} \cos^2\theta_{rf}^\prime ~ L_+
	\right), \label{3.9}
	\end{equation}
	where the index 1 (2) stands for the ordinary (extraordinary) mode,
	$\theta_{rf}$ and $\theta_{rf}^\prime$ are the incident and scattered angles, respectively,
	with respect to the {\it local} magnetic field $\vec B$, as measured in the electron's rest frame,
	$r_0$ is the classical electron radius, and $c$ is the speed of light.  
	The quantities $L_-$ and $L_+$ are the normalized Lorentz profiles, namely
	\begin{equation}
	L_{\pm}= {{\Gamma_{\pm}/2\pi} \over 
		{(\omega_{rf} - \omega^r_{rf})^2 +(\Gamma_{\pm}/2)^2} },
	\label{3.10}
	\end{equation}
	where $\omega_{rf}$ and $\omega^r_{rf}$ are the photon frequency and the resonant
	frequency measured in the electron's rest frame, respectively. 
	The cyclotron line widths (or transition rates)
	$\Gamma_{\pm}$ are given by \citep{Herold1982}
	\begin{equation}
	\Gamma_-={4 \over 3} { {m_ec^2} \over \hbar} \alpha 
	{ {B^2} \over {B_{cr}^2} },
	\label{3.11}
	\end{equation}
	\begin{equation}
	\Gamma_+={2 \over 3} { {m_ec^2} \over \hbar} \alpha 
	{ {B^3} \over {B_{cr}^3} },
	\label{3.12}
	\end{equation}
	where $\alpha$ is the fine structure constant.

    In this work we present polarization-independent calculations and thus we only make use of the polarization averaged differential cross section \citep{Loudas2021}, namely
	\begin{equation}
	{ {d\sigma} \over {d\Omega'} } \approx 
	{ {3\pi r_0c} \over {16} }(1+\cos^2\theta_{rf})(1+\cos^2\theta_{rf}^\prime)
	\left( L_- + {1 \over 2} {B \over {B_{cr}} } L_+ \right).
	\label{3.13}
	\end{equation} 
    
	The resonant energy $E^r_{rf} = \hbar\omega^r_{rf}$, expressed in the electron rest frame, is given by \citep[e.g.,][]{Nobili2008} 
	\begin{equation}
		E^r_{rf} = \dfrac{2E_c}{1 + \sqrt{1 + 2(B/B_{cr})\sin^2\theta_{rf}}}, \label{3.14}
	\end{equation}
	where $E_c=\hbar \omega_c$ is the classical cyclotron energy and $\omega_c = eB/m_e c$. For each scattering event in the code, the new photon energy in the rest frame of the electron after scattering is computed by Eq. (8) of \cite{Loudas2021}.

	\subsection{Formation of the power-law and the scattering feature}
	Some of the photons that are injected isotropically upward escape unscattered, while the rest of them get scattered once or multiple times. It is the latter category of photons, that may criss-cross the shock many times, increasing their energy considerably and eventually producing the high-energy power-law spectrum (see \citealt{Kylafis2014}). However, there is a fraction of those photons, which have energy comparable to the energy difference between the fundamental and the first Landau level (i.e., the cyclotron energy), that undergo resonant scattering with electrons and thus get trapped in the matter until they lose or gain enough energy to escape in the wings. The combination of these processes inevitably leads to the formation of a hard X-ray power-law continuum, including a prominent CRSF in the overall spectrum.

\section{Results} \label{sec4}

    In this section, we present the results of our Monte Carlo polarization-independent calculations in the form of $\der N/\der E$, in arbitrary units, as a function of the photon energy $E$ in units of keV. Every spectrum is normalized so that the integral of $\der N/\der E$ over all energies in the plot is unity. The main parameters in our calculations are:
    \[
    \begin{array}{lp{0.8\linewidth}}
        \tau_\perp & transverse Thomson optical depth,
        \\
        \beta_{ff} & free-fall velocity over the speed of light,
        \\
        B & magnetic-field strength at the shock's altitude,
        \\
        T_e & thermal electrons' post-shock temperature. 
    \end{array}
    \]
    In all cases, the seed photon spectrum is bremsstrahlung (see Sect. \ref{sec3.2}) of temperature equal to the post-shock temperature $T_e$. We study the emergent spectrum under diverse conditions, that is, under different sets of the aforementioned parameters of our model. To achieve high-resolution spectra, we use a minimum of $10^8$ photons in every Monte Carlo run.

    \subsection{Benchmark model}
    \label{sec4.1}
    
    To assess the performance and the validity of our Monte Carlo code and in order to see clearly the formation of a CRSF in the emergent spectrum, we start with a benchmark model. We inject photon beams in the shock, the energy distribution of which is bremsstrahlung of a typical post-shock temperature $kT_e = 10$ keV. We set the local magnetic-field strength to be $B=0.05B_{cr}$, which corresponds to a cyclotron line energy $E_c = 25.55$ keV (typical for accreting NSs), but consider an unrealistic, non-relativistic bulk velocity $v_{ff} = 0.05c$ to suppress the relativistic effects that play a leading role in the spectral formation, as we highlight in Sect. \ref{sec4.2}. We run the code for three different values of the parameter $\tau_\perp$: $0.01,~ 0.1,~ 1$. The low optical depths are nonphysical, but we use them just for this benchmark calculation.  Also, we avoid using optical depths greater than one, as we aim only to test the effects of resonant scattering at this stage. 

       \begin{figure}
        \includegraphics[width=9cm]{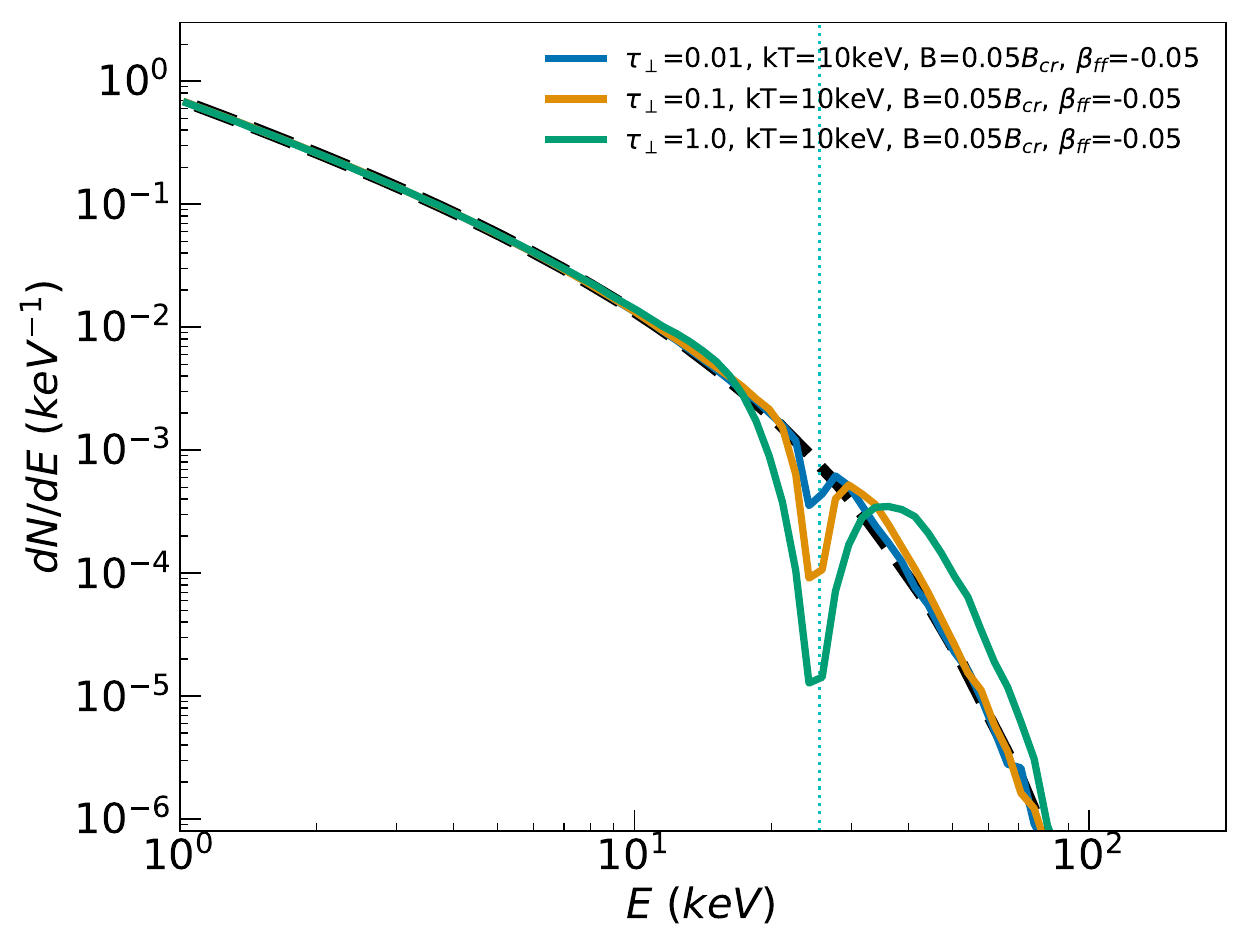}
        \caption{
            Angle- and polarization-averaged emergent spectrum $\der N/\der E$ as a function of energy $E$ for various transverse Thomson optical depths. The bulk velocity was set to $v_{ff}=0.05c$. The black dashed line represents the seed Bremsstrahlung photon spectrum of temperature $kT_e = 10$ keV. The blue solid line stands for $\tau_\perp = 0.01$, the golden solid line is for $\tau_\perp = 0.1$, and the green solid line refers to $\tau_\perp = 1$. The vertical blue dotted line indicates the classical cyclotron line energy $E_{c} = 25.55$ keV. All spectra are normalized to unity.
        }
        \label{fig1} 
    \end{figure} 
    
    In Fig. \ref{fig1}, we show the results. The black dashed line represents the input bremsstrahlung spectrum, while the rest of them correspond to the resulting angle- and polarization-averaged spectra of our MC simulations for different values of the transverse Thomson optical depth: $\tau_\perp = 0.01$ (blue), $\tau_\perp = 0.1$ (golden), $\tau_\perp = 1$ (green).  For visualization purposes, we have added a vertical light-blue dotted line that indicates the value of the classical cyclotron energy used in these runs. Although the continuum is not altered noticeably, a prominent absorption feature arises around the classical cyclotron energy and becomes broader and deeper as the transverse optical depth increases. Even for $\tau_\perp = 0.1$ (golden line), which refers to an optically thin medium for the continuum, the absorption feature is clearly formed in the spectrum. In addition, bump features appear in the wings, with the one in the right wing, that is, for $E \gtrsim E_{c}$, being more prominent. 
    
    The emergent spectra, shown in Fig. \ref{fig1}, exhibit the anticipated characteristics. In particular, a prominent absorption feature appears near the cyclotron energy, which reflects the fact that the mean free path for photons of energy comparable to the resonant energy is very small (see Sect. \ref{sec3.4}), though the medium is optically thin to Thomson scattering. Essentially, a considerable fraction of these photons undergo resonant scattering and may criss-cross the shock (up and down) multiple times until their energies become significantly higher or lower than the resonant one and they acquire a large mean free path and escape the column sideways. These photons appear in the absorption feature's left or right wing as bump features. The emergence of bumps in the neighborhood of the absorption features is a natural consequence of the shock's medium being pure scattering, which dictates that the total number of photons must be conserved.  
    
    Regarding the shape of the absorption feature, we notice that as the transverse Thomson optical depth increases, more and more photons of energies around the resonant one get resonantly scattered, resulting in a deeper, as well as broader absorption feature. This trend is clearly depicted in Fig. \ref{fig1}. On the other hand, whether most of the resonantly scattered photons escape in the left wing or in the right one is determined mainly by the post-shock temperature. In fact, every time a photon enters the post-shock region, it is more likely to experience resonant scattering, since the cross-section for propagation of photons in a thermal plasma is broadened due to the thermal motion of the electrons (see Appendix \ref{App.A}), and hence, photons are  scattered by the thermal plasma until they gain or lose enough energy to escape. For low-temperature plasma (cold), photons lose energy, on average, and escape primarily in the left wing, while for high-temperature (hot) plasma, photons gain energy, on average, and escape primarily in the right wing (see also Sect. \ref{sec4.3}).     

    The results of the benchmark calculations shown in Fig. \ref{fig1} enable us to understand the main physical processes, but they are not appropriate for describing shocks in accretion columns of NSs, because we have considered a non-relativistic bulk velocity above the shock. For a typical accreting NS, the free-fall velocity would be something like $v_{ff}=0.6c$.

    \begin{figure}
        \includegraphics[width=9cm]{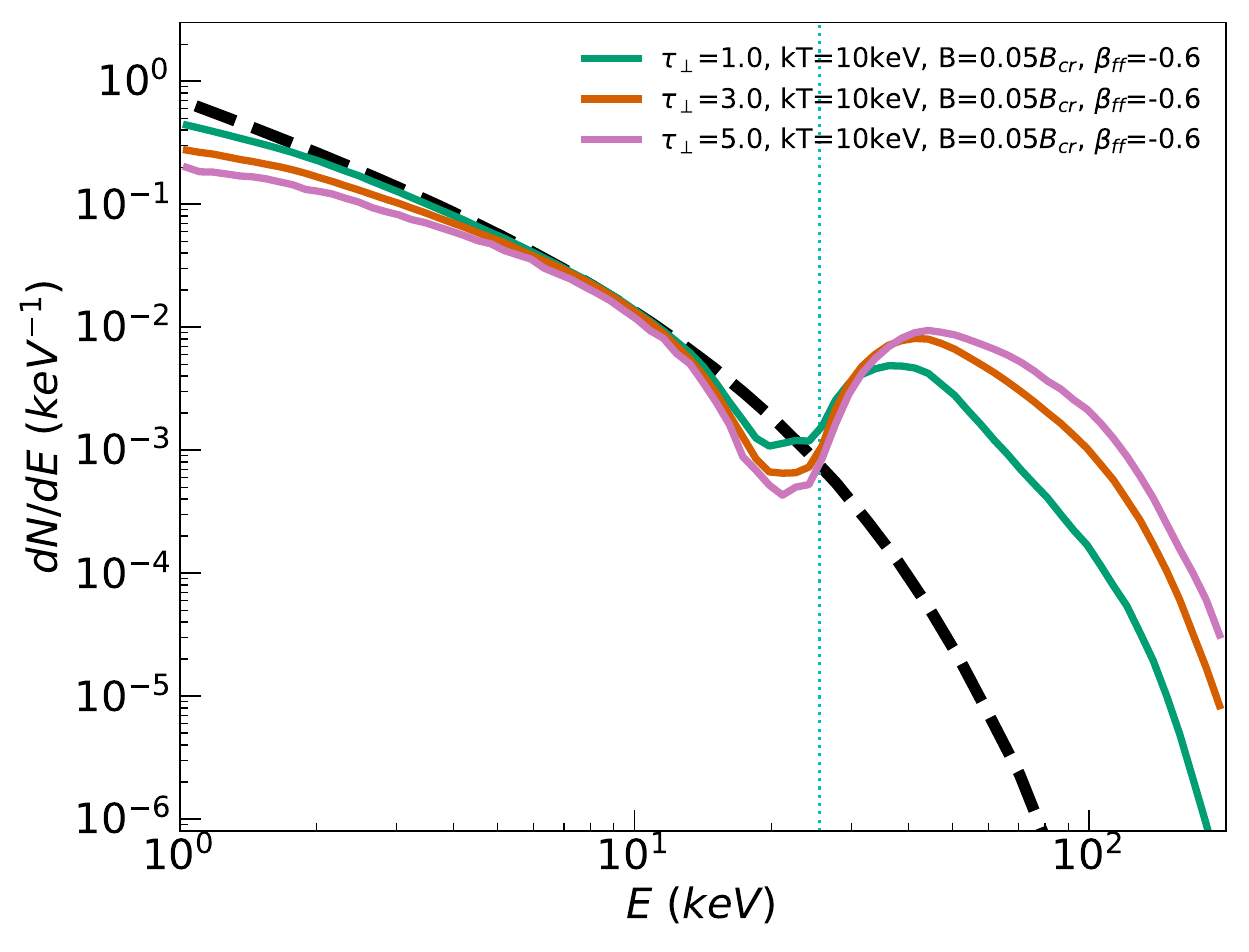}
        \caption{
             Angle- and polarization-averaged emergent spectrum $\der N/\der E$ as a function of energy $E$ for various transverse Thomson optical depths. The bulk velocity was set to $v_{ff}=0.6c$. The black dashed line represents the seed Bremsstrahlung photon spectrum of temperature $kT_e = 10$ keV. The green solid line refers to $\tau_\perp = 1$, the orange solid line accounts for $\tau_\perp = 3$, and the purple solid line is for $\tau_\perp = 5$. The vertical blue dotted line indicates the classical cyclotron line energy $E_{c} = 25.55$ keV. All spectra are normalized to unity.  
        }
        \label{fig2} 
    \end{figure}

	\subsection{Effects of the bulk-motion} \label{sec4.2}

    As we discussed in Sect. \ref{sec2}, the effect of the bulk-motion Comptonization is of unique importance in shaping the resulting spectrum, as the accretion flow gets decelerated in the shock supplying its accretional energy to photons, thereby altering their energy distribution. In order to investigate the effect of the relativistic bulk motion of the accreting plasma (pre-shock region), we run the MC code for $\tau_{\perp} = 1$, 3, and 5 with a bulk velocity, which is now chosen to be $v_{ff} = 0.6c$, appropriate for a NS of mass $M = 1.4 ~\mathrm{M_\odot}$ and radius $R = 10~\mathrm{km}$.  
    
    In Fig. \ref{fig2}, we present the results of our MC simulations for different values of the transverse Thomson optical depth, $kT_e = 10$ keV, and $E_c = 25.55$ keV. The input spectrum is displayed with a black dashed line, while the emergent spectra are displayed with colored solid lines: $\tau_\perp = 1$ (green), $\tau_\perp = 3$ (orange), and $\tau_\perp = 5$ (purple). The vertical light-blue dotted line indicates the classical cyclotron energy. As expected, a prominent absorption feature arises in the spectrum. However, in this case, the spectra differ significantly from those presented in Sect. \ref{sec4.1}. First, the centroid of the absorption feature is shifted to lower energies by $\sim 30\%$ with respect to the classical cyclotron line, and there is a hint that the shift slightly decreases with increasing $\tau_\perp$, though higher precision and fitting the whole spectrum would be required for trustful evidence of this trend. The depth of the absorption feature increases with increasing $\tau_\perp$, in line with Sect. \ref{sec4.1}. However, the absorption feature's width is significantly broader than the one shown in the results of the benchmark model discussed in Sect. \ref{sec4.1}. Third, a bump appears in the right wing of the line, followed by a hard X-ray tail, similar to the reported spectra of accreting XRPs (e.g., V 0332+53: \citealt{Doroshenko2017}; 4U 1538-522: \citealt{Sharma2023}). In addition, a power-law-like spectrum arises at low energies and extends up to the CRSF. The power-law index decreases, that is the spectrum becomes harder, as the transverse optical depth increases, in line with \cite{Kylafis2014}.

    The physical explanation of the characteristics of the emergent spectrum  follows. As we discussed in Sect. \ref{sec4.1}, the resonant cross-section implies an extremely small mean free path for the propagation of photons of energy comparable to the resonant energy. However, one should take into account that photons emitted just below the shock (shock/lab frame) appear blue-shifted in the in-falling plasma (bulk frame) as a consequence of relativity (recall that $v_{ff} = 0.6 c$). As a result, photons that have energies smaller than the classical cyclotron energy, depending on their direction of propagation, will undergo resonant scattering with the free-falling electrons and subsequently will appear in the blue (right) wing of the CRSF, on average, due to relativistic beaming (i.e., an inverse Lorentz transformation from the bulk-frame back to the shock-frame).
    Therefore, photons that get resonantly scattered from the in-falling electrons are Doppler boosted twice, gaining, in this way, a considerable amount of energy, on average.   
    This discussion explains both the shifting of the CRSF to lower energies than the classical cyclotron one, due to the Doppler boosting effect, and the creation of a bump in the right wing of the absorption feature with a hard X-ray tail, because scattering preserves the total number of photons.  Notice that the resonantly upscattered photons acquire energies {\it significantly larger} than bremsstrahlung allows.
    Since photons are emitted isotropically upwards from below the shock, photons of different directions of propagation (with respect to the magnetic field axis) will experience a different Doppler boosting from the shock frame to the bulk frame; thus, for resonant scattering between photons and free-falling electrons, a range of different photon energies is sampled in view of multiple directions of photon propagation. This leads to the formation of a broad absorption feature, though shifted to lower energies due to the Doppler effect, as it can be seen in Fig. \ref{fig2}. 
    
    As the transverse optical depth increases, a larger fraction of photons undergo resonant scattering and criss-cross the shock; thus, the CRSF becomes more prominent. But, in the post-shock region, the electrons are thermal with negligible bulk velocity, and as a result, the absorption feature that is formed below the shock by the photons that enter this region is almost centered at the classical cyclotron energy, though thermally broadened due to the thermally averaged cross-section (App. \ref{App.A}). It turns out that the combination of the Doppler boosting (pre-shock region) and the thermal Doppler broadening (post-shock region) result in a wide, prominent CRSF that is shifted towards smaller energies than the classical cyclotron energy.
    
    Finally, the continuum on the left of the CRSF becomes harder as the $\tau_\perp$ increases and exhibits a power-law-like shape as a natural consequence of the first-order Fermi energization (for a thorough discussion see \citealt{Kylafis2014}).

    \subsection{Effects of the post-shock temperature} \label{sec4.3}


     \begin{figure}
        \includegraphics[width=9cm]{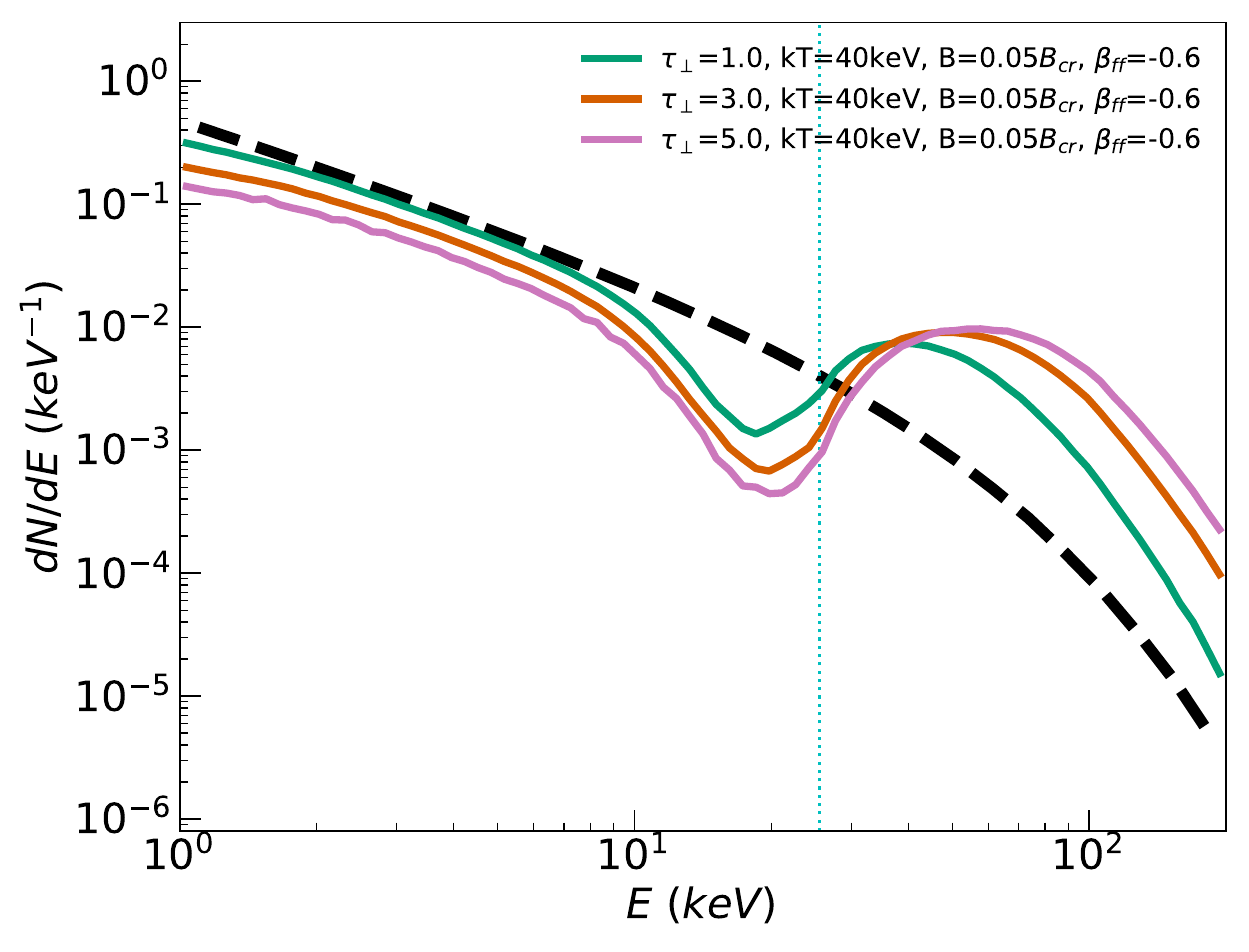}
        \caption{
            Same as in Fig. \ref{fig2}, but for $kT_e = 40$ keV.  
        }
        \label{fig3} 
    \end{figure}

     \begin{figure}
        \includegraphics[width=9cm]{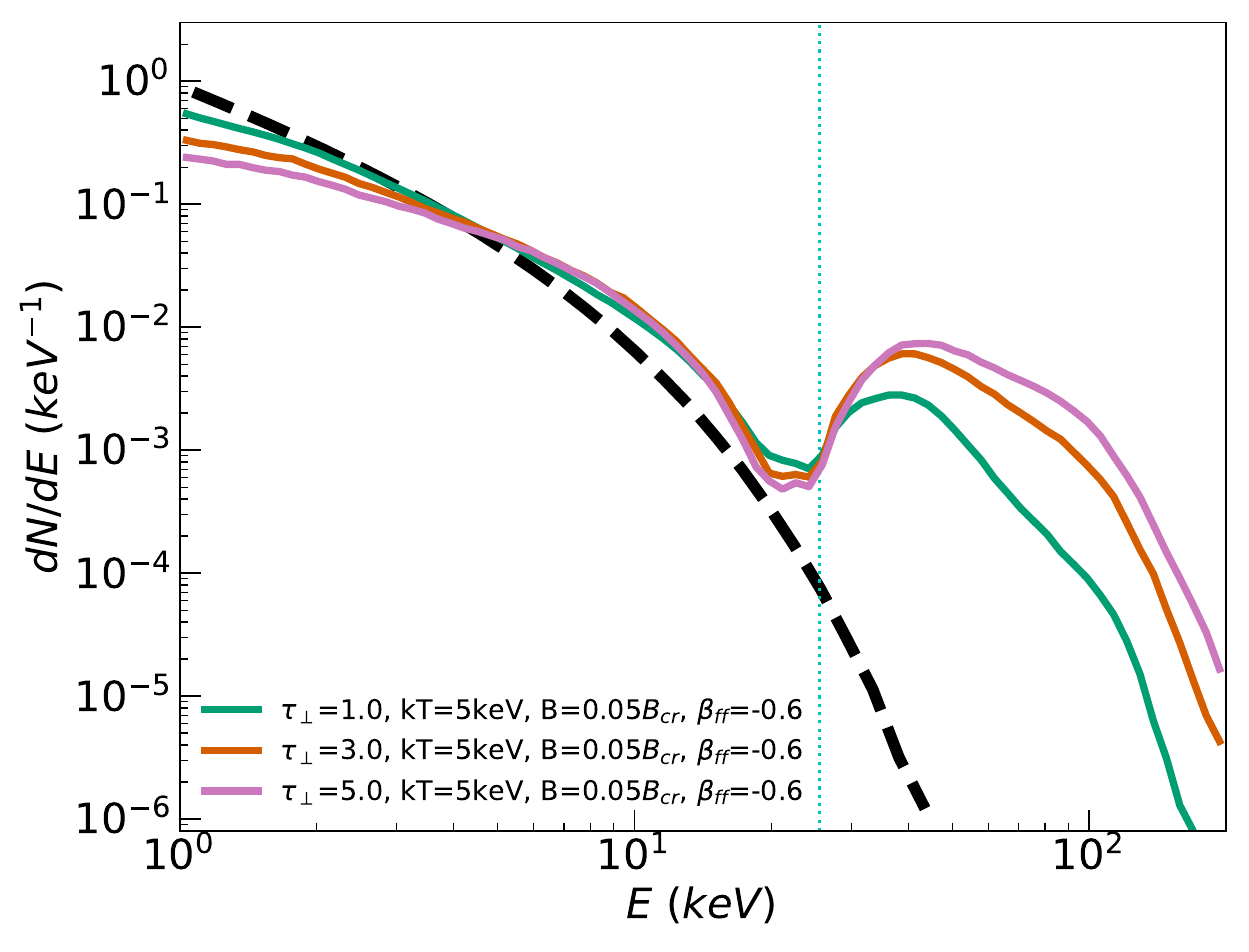}
        \caption{
            Same as in Fig. \ref{fig2}, but for $kT_e = 5$ keV.  
        }
        \label{fig4} 
    \end{figure}

    As we argued in Sect. \ref{sec2}, the post-shock temperature is unknown, although it is related to the accretional energy of the electrons. For this reason, we run our MC code for a wide range of $kT_e$ values to study its effect on the emergent spectrum.

    Figure \ref{fig3} is similar to Fig. \ref{sec2}, but for $kT_e = 40$ keV. The characteristics of the spectra are the same as the ones shown in Fig. \ref{fig2}. Again, a prominently-shifted absorption line is formed, accompanied by a bump in the right wing and a hard X-ray tail. The shift of the line centroid again seems to slightly decrease with increasing $\tau_\perp$. This trend, although not-well established, could likely be a consequence of the interplay between the bulk-motion resonant scattering in the pre-shock region and the thermal resonant scattering in the post-shock one. In fact, for larger $\tau_\perp$, a greater fraction of photons criss-crosses the shock, which implies that the number of photons that are resonantly scattered both in the pre-shock region and in the post-shock one increases, leading, in this way, to the overlapping of the two different resonant scattering processes. This, in turn, could reduce the effect of the bulk motion resonant scattering, especially when the post-shock temperature is high, and eventually, the CRSF may be only partially shifted to the left.
    
\begin{figure}
        \includegraphics[width=9cm]{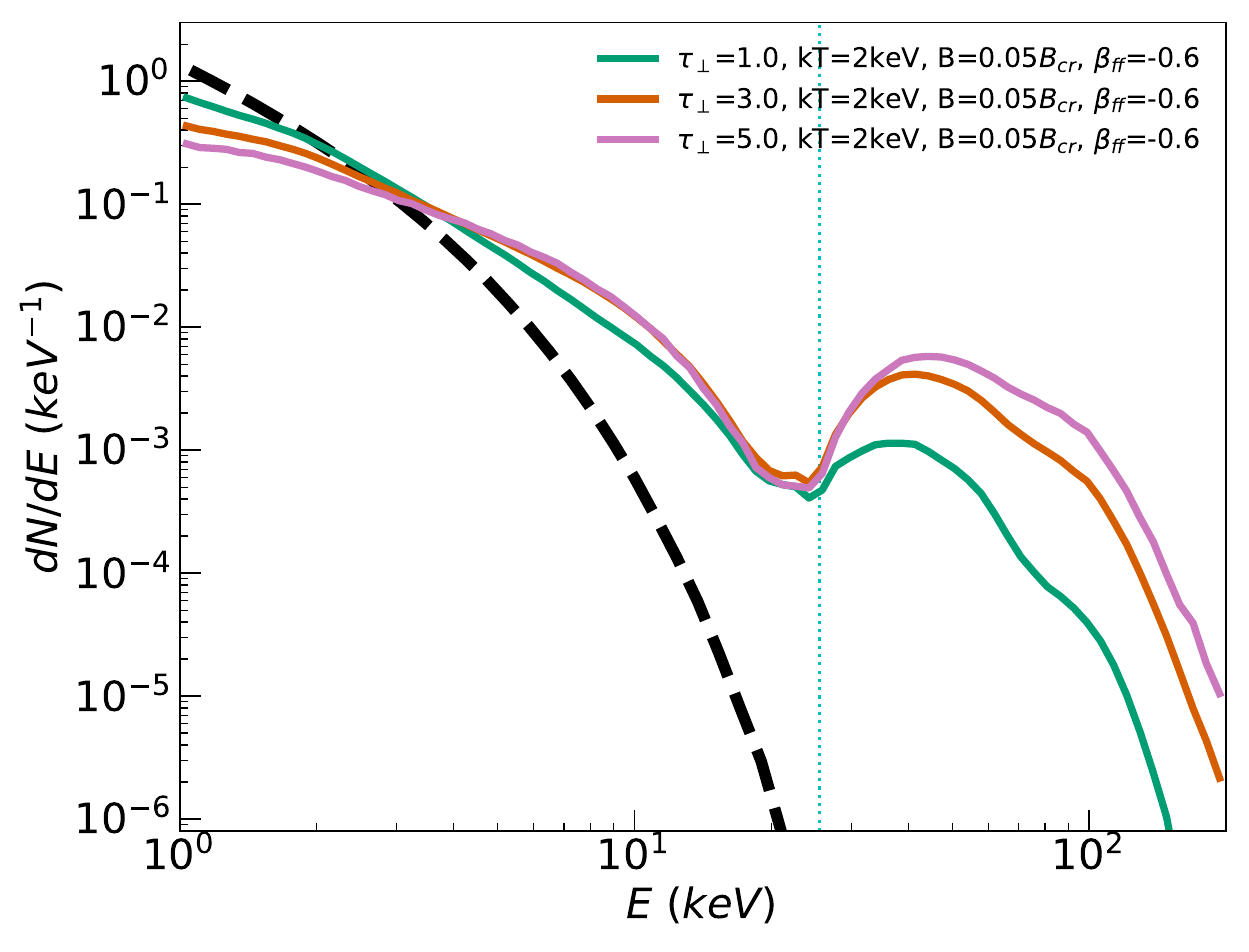}
        \caption{
            Same as in Fig. \ref{fig2}, but for $kT_e = 2$ keV. 
        }
        \label{fig5} 
    \end{figure}

    However, the width and the depth of the CRSF are somewhat greater than those in Fig. \ref{fig2} for the same value of $\tau_\perp$. The 
    broadening and the deepening of the absorption feature with the post-shock temperature are consequences of two factors: 1) the seed photon spectrum around the resonant energy is flatter for greater temperatures, thereby allowing for a larger number of photons to be resonantly scattered, and 2) the higher the $kT_e$, the broader the thermally averaged resonant cross-section (see Appendix \ref{App.A}), and thus the wider the absorption feature generated in the post-shock region. The combination of these reasons and the preservation of the total number of photons explains the fact that the peak of the bump (in the right wing) appears at relatively higher energies and is stronger and broader for larger post-shock temperatures with respect to the ones in Fig. \ref{fig2} for the same $\tau_\perp$. Figures \ref{fig4} \& \ref{fig5} are similar to Fig. \ref{sec2}, but for $kT_e = 5$ keV and $kT_e = 2$ keV, respectively. The conclusions are the same.

\begin{figure}
        \includegraphics[width=9cm]{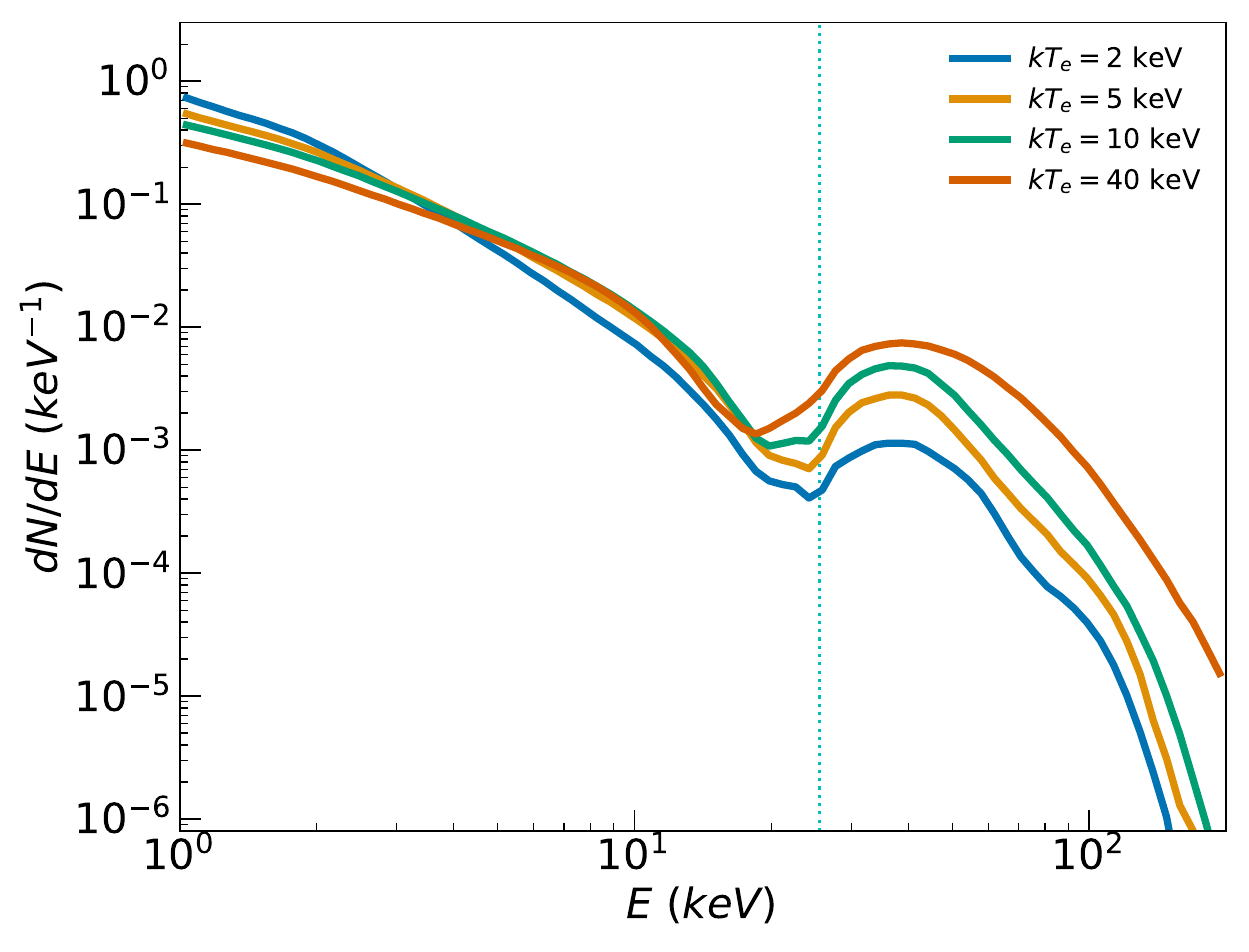}
        \caption{
            Angle- and polarization-averaged emergent spectrum $\der N/\der E$ as a function of energy $E$ for $\tau_\perp = 1$, $v_{ff}=0.6c$, and $E_{c} = 25.55$ keV. Solid lines correspond to different post-shock temperatures $kT_e$: $2 ~\mathrm{keV}$ (blue); $5 ~\mathrm{keV}$ (golden); $10 ~\mathrm{keV}$ (green); and $40 ~\mathrm{keV}$ (orange). Vertical blue dotted line indicates the classical cyclotron line energy. All spectra are normalized to unity.  
        }
        \label{fig6} 
    \end{figure}

\begin{figure}
        \includegraphics[width=9cm]{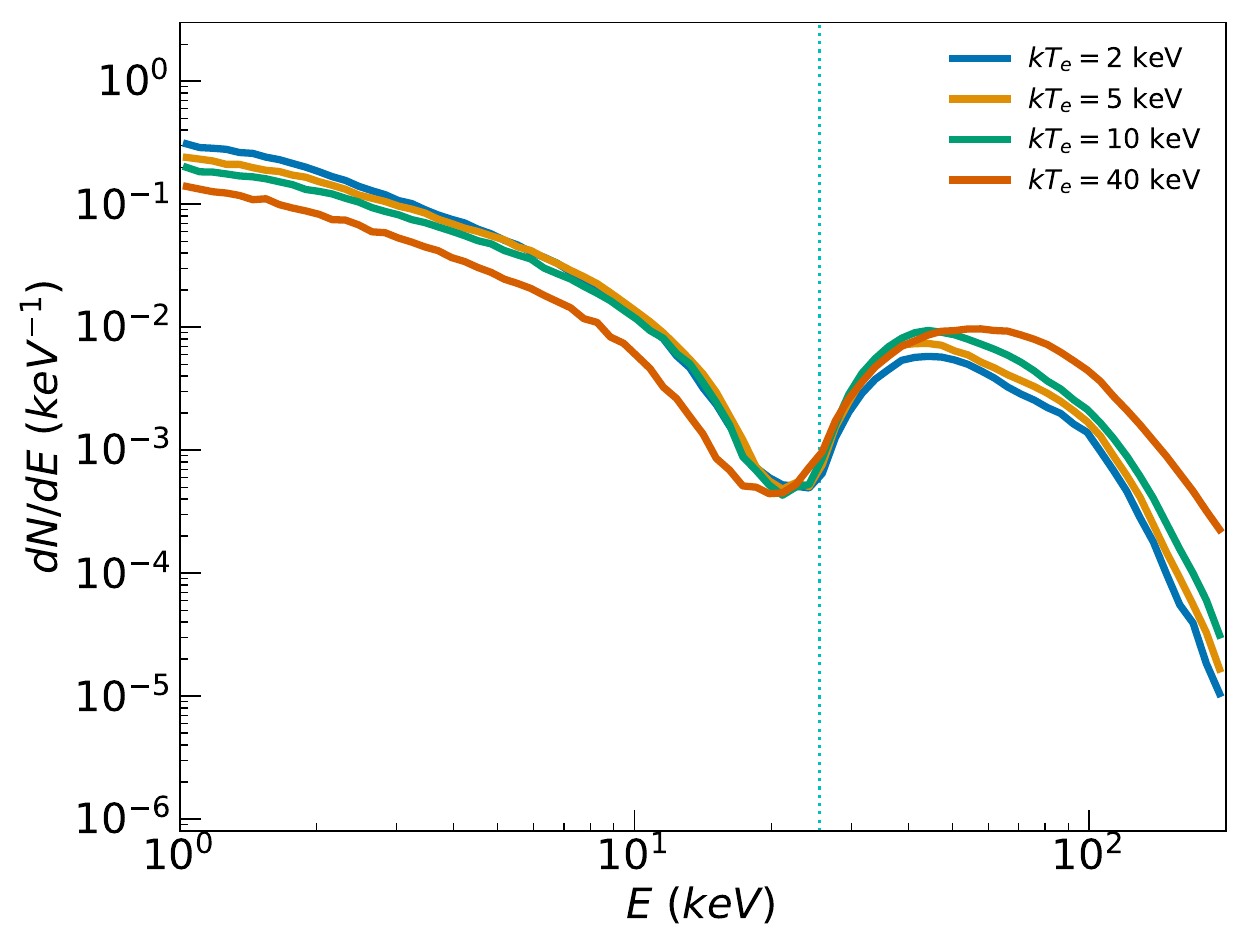}
        \caption{
            Same as in Fig. \ref{fig6}, but for $\tau_\perp = 5$. 
        }
        \label{fig7} 
    \end{figure}

    For completeness, and for direct comparison, we plot in Fig. \ref{fig6} the emergent spectrum for $B=0.05~B_{cr}$, $v_{ff} = 0.6 c$, and $\tau_\perp = 1$ for four different values of the post-shock temperature: $kT_e = 2$ keV (blue), $kT_e = 5$ keV (golden), $kT_e = 10$ keV (green), $kT_e = 40$ keV (orange). Figure \ref{fig7} is the same as Fig. \ref{fig6}, but for $\tau_\perp = 5$. 
    The two figures indicate that the post-shock temperature, as well as the transverse optical depth, affect the shape of the CRSF. The shift of the CRSF is greater for larger temperatures and relatively small optical depths ($\tau_\perp \sim 1$). This is because the spectrum is flatter for higher temperatures, and a larger fraction of the seed photons undergo resonant scattering with the in-falling electrons, leading to stronger (shifted-) absorption and bump features (see Fig. \ref{fig6}). On the other hand, the effect of thermal resonant scattering (post-shock region) is smeared out by the strong bump feature produced via bulk-resonant scattering and hence it becomes less efficient when it comes to reducing the shift of the line produced by the bulk resonant scattering as long as the transverse optical depth is small (see Fig. \ref{fig6}). However, as $\tau_\perp$ increases, more and more photons are scattered downwards by the accretion flow, crossing the shock and entering the post-shock region. Therefore, a greater fraction of photons get thermal resonantly scattered below the shock leading to a stronger, non-shifted absorption feature, which impacts to a greater extent the overall shape of the emergent absorption feature and the right wing. This is clearly inferred by comparing Fig. \ref{fig7} to Fig. \ref{fig6}.   

    \subsection{Effects of the transverse optical depth} \label{sec4.4}

    As we discussed in the previous sections, the transverse Thomson optical depth has a dominant role in forming the overall spectrum. Its effects are summarized below. First, the depth of the absorption feature increases remarkably, i.e., it becomes more prominent for higher values of $\tau_\perp$. Second, the larger the optical depth, the smaller the shift of the line centroid to lower energies seems to be, although resolution limits and noise prevent us from obtaining conclusive evidence of this point. Third, the bump in the right wing becomes stronger with increasing $\tau_\perp$. Fourth, the continuum becomes harder and the power-law at energies lower than the cyclotron energy becomes flatter (i.e., the power-law index decreases) for higher values of $\tau_\perp$. Fifth, the hard X-ray tail, that emerges on the right of the CRSF, extends to higher energies as the optical depth increases and reaches energies significantly larger than the corresponding temperature dictates.

    \subsection{Effects of the magnetic-field strength}
	\label{sec4.5}

\begin{figure}
        \includegraphics[width=9cm]{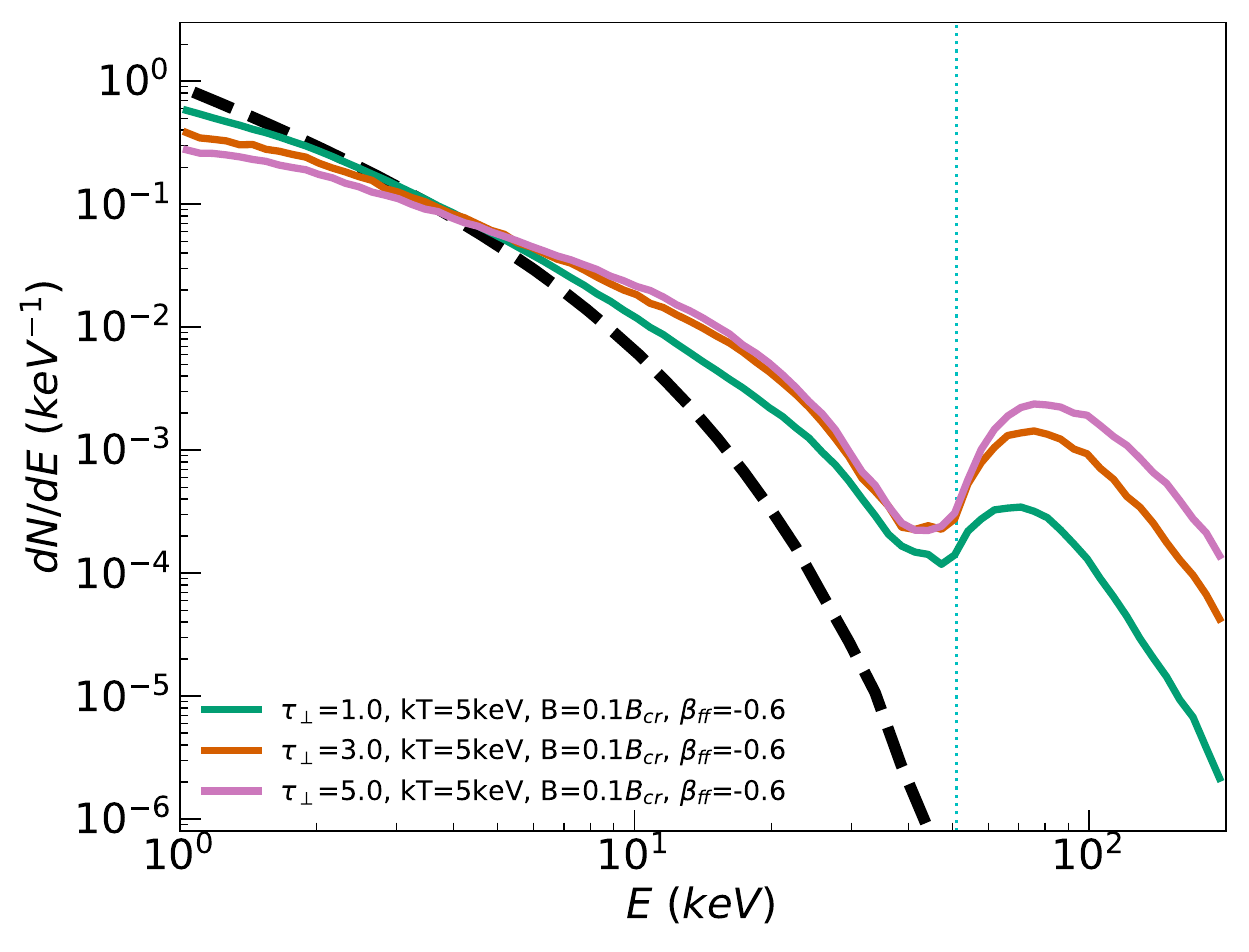}
        \caption{
            Same as in Fig. \ref{fig4}, but for $B=0.1 ~\mathrm{B_{cr}}$ (or $E_c = 51$ keV).  
        }
        \label{fig8} 
    \end{figure}

    As we discussed in Sect. \ref{sec3.4}, the resonant energy is approximately proportional to the magnetic-field strength. Thus, the emergent CRSF scales almost linearly with $B$ and it will appear near the classical cyclotron energy, though shifted to lower energies, as we demonstrated in Sect. \ref{sec4.2}. Although the magnetic-field strength on the surface of an accreting magnetic NS spans from $\sim 10^{12}~\mathrm{G}$ to $\sim 10^{13}~\mathrm{G}$, we have chosen in this work to present results for a typical magnetic field strength that corresponds to $E_{c} = 25.55~\mathrm{keV}$, i.e., $B/B_{cr} = 0.05$. For demonstration purposes, we run our MC code for the set of parameters used in Fig. \ref{fig4}, but for $B/B_{cr} = 0.1$, which corresponds to $E_c = 51$ keV. We present the results in Fig. \ref{fig8}. The black solid line is the input spectrum, while the rest refer to the emergent spectra for different values of the transverse optical depth: $\tau_\perp = 1$ (green), $\tau_\perp = 3$ (orange), and $\tau_\perp = 5$ (purple). The conclusions are identical to those made in the previous sections. As expected, the CRSF appears near the resonant energy, though redshifted. 

    We note that the magnetic-field strength varies significantly along the column under the dipolar magnetic-field assumption. Therefore, the emergent CRSF's centroid is expected to vary with the shock's height. In fact, as the shock's height $H$ increases, the B-field strength drops as $(1+H/R)^{-3}$, and hence one might consider that the rate of change of the observed cyclotron line energy with the shock's height is dominated by this drop of the magnetic-field strength along the column. However, as we demonstrated in the previous sections, the bulk-motion speed also plays a crucial role in the line-forming process. In particular, it shifts the line centroid to lower energies and the amount of shifting depends on the free-fall speed at height $H$. Given that the free-fall velocity drops as the shock's height increases, the shift due to the bulk resonant scattering decreases and thus affects the correlation between the line centroid and the shock's height (i.e., the X-ray luminosity). Therefore, the red-shifting effect must be considered when one tries to explain the variation of the cyclotron line energy with the X-ray luminosity. However, a further discussion on this topic is out of this paper's scope and it will be addressed thoroughly in a follow-up study.

    \section{Discussion \& conclusions} \label{sec5}
	
	In this study, we explored the possibility of cyclotron-line formation in the radiative shock in the accretion column of an accreting magnetic neutron star of super-critical X-ray luminosity, that was first proposed by \cite{Basko1976}. We developed a Monte Carlo code, implementing a fully relativistic numerical scheme to carry out radiative-transfer calculations in the shock and compute the emergent spectrum. We used a model in which the shock is a pure mathematical discontinuity in a cylindrical column, assuming that above the shock, matter falls freely, while below it, the plasma is thermal, having negligible flow velocity. We assumed a bremsstrahlung seed photon distribution. The main parameters of our model are the transverse Thomson optical depth, the local magnetic-field strength at the shock's altitude, the post-shock electron's temperature, and the free-fall velocity at the shock's altitude. We have properly treated both the bulk-resonant scattering by free-falling electrons in the pre-shock region and the thermal-resonant scattering by thermal electrons in the post-shock one, employing an accurate prescription for resonant scattering in accreting NSs \citep{Loudas2021}. 

    We provided insights into whether cyclotron line formation in the radiative shock is possible and showed that prominent CRSFs are generated in the radiative shock and appear in the overall spectrum. We found that the characteristics of the emergent spectrum are: 1) a power-law at energies below the energy of the CRSF; 2) a deep, broad, and shifted to lower energies absorption feature; 3) a bump in the right wing of the CRSF; and 4) a hard X-ray tail that extends above $\sim 100$ keV. We remark that these features are generally observed in accreting NSs (e.g., V 0332+53; \citealt{Doroshenko2017}), though fitting data is still premature. We would like to make clear that the part of the spectrum next to the funtamental CRSF (blue wing) could be altered by the redistribution of photons undergoing resonant scattering due to harmonics, leading to a sequence of approximately evenly spaced cyclotron lines (corresponding to harmonics) imprinted in the hard X-ray tail of the spectrum. This has already been observed in V 0332+53 (\citealt{Tsygankov2006}), but such a calculation is out of the scope of this study.
    
    The power-law spectrum that is formed at low energies is a natural consequence of the combination of bulk- and thermal-Comptonization, and is due to the first-order Fermi energization. The spectral power-law index depends both on the transverse Thomson optical depth and the post-shock temperature. The larger the optical depth, the flatter the spectrum. Similarly, the higher the post-shock temperature, the lower the power-law index. The power-law-like feature extends up to the CRSF.  Thus, its extent depends on the magnetic field strength at the shock. 
    
    We demonstrated that the absorption feature's depth, shift, and width are mainly affected by the free-fall velocity, the transverse optical depth, and the post-shock temperature. The absorption feature becomes deeper and broader as the optical depth or the post-shock temperature increases. The line centroid is shifted to lower energies due to the bulk resonant scattering, in line with the findings of \cite{Nishimura2014}. The amount of shift is greater for higher values of the free-fall velocity. For $v_{ff} = 0.6c$, typical for accreting NSs, the CRSF is shifted by $\sim(20-30)\%$ and the shift seems to slightly decrease as the optical depth increases (e.g., Fig. \ref{fig3}), although this hint needs to be tested by increasing the resolution in the simulations and fitting accurately the line profile. This result implies that for precise measurements of the magnetic field strength through cyclotron line observations, this effect must be quantified and taken into account. 

    The bump always appears in the right wing and is followed by a hard X-ray tail. The main parameters affecting these features' shape are the free-fall velocity, the optical depth, and the post-shock temperature. The bump is mainly formed by the upscattering of photons that undergo resonant scattering with the in-falling electrons and is reprocessed by the thermal electrons in the post-shock region. Its strength and width increase and the hard X-ray tail extends to higher energies for greater optical depth or higher post-shock temperature values. Its peak energy depends primarily on the free-fall velocity through the Doppler effect, but it also increases significantly with the optical depth and the post-shock temperature.         

    It should be noted that the derived spectra, although similar, are generally harder than observed, presumably indicating less bulk Comptonization of the emergent photons, but the overall conclusion depends also on the contribution of the emission of the accretion column below the shock. If one takes into account in the emergent spectrum the radiation emitted in the accretion column far below the shock, provided the emission profile along the column is known, and assumes that $kT_e < E_c$, then the total emergent spectrum from the accretion column may be softer than the one produced solely at the radiative shock.

    One source of uncertainty in our calculations arises from the fact that we considered a rather simplified picture for the accretion column/shock structure. We assumed the shock to be: 1) stationary, 2) plane-parallel, and 3) of infinitesimal width (i.e., a mathematical discontinuity), satisfying the standard Rankine-Hugoniot jump conditions for a radiation-dominated shock of extremely high Mach number far upstream (appropriate for accreting XRPs in the super-critical luminosity regime). Such assumptions result in a velocity drop of $(\approx1/7)$ across the shock \citep{Basko1976}. Recent and more advanced studies on the topic have revealed a more complex picture though (e.g., \citealt{Postnov2015}). \cite{Zhang2022} have shown that the shock front exhibits high-frequency oscillations, has paraboloidal-like shape, and is of finite size (i.e., continuous) rather than a mathematical discontinuity. How these findings would alter our conclusions is difficult to determine without performing more detailed MC radiative transfer simulations. One scenario would be that the Fermi energization process of soft photons at the shock may not be that efficient due to the larger spatial scale of the shock transition (i.e., a smoother density and velocity gradient across the shock), resulting in an overall weaker energization. Apparently, this calculation, although important, far exceeds the scope of this paper.

    Another caveat in our consideration is related to the choice of the seed photon energy spectrum. Although the plasma in the post-shock temperature is strongly magnetized, we chose to employ a field-free Bremsstrahlung emission spectrum below the shock. A more rigorous choice would be to consider magnetic Bremsstrahlung photons that have undergone thermal Comptonization in the post-shock region. However, in this study, we focused on studying the first-order Fermi energization of soft photons in the shock accounting for the effects of bulk as well as thermal Comptonization, and resonant scattering, but ignoring the detailed shape of the seed photon distribution as minor changes in the seed photon spectrum would not alter the results, provided the shock resides in the optically thick regime.
    
    We note that magnetic Bremsstrahlung emission (e.g., \citealt{Riffert1999}) induces anisotropy and yields an angle-dependent seed photon emission spectrum, while cyclotron resonant scattering leads to an angle-dependent emission line in the magnetic Bremsstrahlung spectrum. These differences may affect the subtle properties of the line features obtained in this work, but would not alter the conclusions, as the thermally- and angle-averaged emission spectrum of magnetic Bremsstrahlung agrees fairly well with the field-free one, so long as the photon energy is smaller than the cyclotron energy (see Sect. 7.3 in \citealt{Becker2007}). Given that in most cases in our study the temperature is smaller than the cyclotron energy (except when $kTe=40 ~\mathrm{keV} > E_c=25.5 ~\mathrm{keV}$),
    the choice between those seed photon distributions would not affect significantly the results.
    
    In summary, we presented for the first time spectral formation in a radiative shock in the accretion column of an accreting magnetic neutron star, giving special emphasis on the formation of the cyclotron resonant scattering feature. We offered a qualitative description of the spectral formation, we provided a quantitative demonstration of the characteristics of the emergent spectra, and studied their dependence on the conditions encountered in the radiative shock. Radiation Magneto-Hydrodynamic (RMHD) simulations (e.g., \citealt{Zhang2022}) in accretion column-like structures or hollow funnel geometries, allowing for resonant scattering that enhances matter-radiation interaction, could be extremely beneficial in advancing cyclotron line formation calculations in the future. 
	
	\begin{acknowledgements}
		We would like to thank the anonymous referee for providing a constructive review with insightful comments and important suggestions that helped us improve this manuscript and offered new ideas to explore in the future. 
		NL acknowledges support by the European Research Council (ERC) under
		the HORIZON ERC Grants 2021 programme under grant agreement No. 101040021. NL acknowledges the "Summer School for Astrostatistics in Crete" for providing training on the statistical methods adopted in this work.
	\end{acknowledgements}

	\bibliography{references}

\begin{thebibliography}{54}
\expandafter\ifx\csname natexlab\endcsname\relax\def\natexlab#1{#1}\fi

\bibitem[{{Basko} \& {Sunyaev}(1975)}]{Basko1975}
{Basko}, M.~M. \& {Sunyaev}, R.~A. 1975, \aap, 42, 311

\bibitem[{{Basko} \& {Sunyaev}(1976)}]{Basko1976}
{Basko}, M.~M. \& {Sunyaev}, R.~A. 1976, \mnras, 175, 395

\bibitem[{{Becker} {et~al.}(2012){Becker}, {Klochkov}, {Sch{\"o}nherr},
  {Nishimura}, {Ferrigno}, {Caballero}, {Kretschmar}, {Wolff}, {Wilms}, \&
  {Staubert}}]{Becker2012}
{Becker}, P.~A., {Klochkov}, D., {Sch{\"o}nherr}, G., {et~al.} 2012, \aap, 544,
  A123

\bibitem[{{Becker} \& {Wolff}(2007)}]{Becker2007}
{Becker}, P.~A. \& {Wolff}, M.~T. 2007, \apj, 654, 435

\bibitem[{{Becker} \& {Wolff}(2022)}]{Becker2022}
{Becker}, P.~A. \& {Wolff}, M.~T. 2022, \apj, 939, 67

\bibitem[{{Bisnovatyi-Kogan} \& {Fridman}(1970)}]{Bisnovatyi1970}
{Bisnovatyi-Kogan}, G.~S. \& {Fridman}, A.~M. 1970, \sovast, 13, 566

\bibitem[{{Bykov} \& {Krasil'Shchikov}(2004)}]{Bykov2004}
{Bykov}, A.~M. \& {Krasil'Shchikov}, A.~M. 2004, Astronomy Letters, 30, 309

\bibitem[{Cashwell \& Everett(1959)}]{Cashwell1959}
Cashwell, E. \& Everett, C. 1959, A Practical Manual on the Monte Carlo Method
  for Random Walk Problems, International tracts in computer science and
  technology and their application (Pergamon Press)

\bibitem[{{Chen} {et~al.}(2021){Chen}, {Wang}, {Tang}, {Ding}, {Tuo},
  {Mushtukov}, {Nishimura}, {Zhang}, {Ge}, {Song}, {Lu}, {Zhang}, \&
  {Qu}}]{Chen2021}
{Chen}, X., {Wang}, W., {Tang}, Y.~M., {et~al.} 2021, \apj, 919, 33

\bibitem[{{Doroshenko} {et~al.}(2017){Doroshenko}, {Tsygankov}, {Mushtukov},
  {Lutovinov}, {Santangelo}, {Suleimanov}, \& {Poutanen}}]{Doroshenko2017}
{Doroshenko}, V., {Tsygankov}, S.~S., {Mushtukov}, A.~A., {et~al.} 2017,
  \mnras, 466, 2143

\bibitem[{{F{\"u}rst} {et~al.}(2018){F{\"u}rst}, {Falkner}, {Marcu-Cheatham},
  {Grefenstette}, {Tomsick}, {Pottschmidt}, {Walton}, {Natalucci}, \&
  {Kretschmar}}]{Fuerst2018}
{F{\"u}rst}, F., {Falkner}, S., {Marcu-Cheatham}, D., {et~al.} 2018, \aap, 620,
  A153

\bibitem[{{F{\"u}rst} {et~al.}(2014){F{\"u}rst}, {Pottschmidt}, {Wilms},
  {Tomsick}, {Bachetti}, {Boggs}, {Christensen}, {Craig}, {Grefenstette},
  {Hailey}, {Harrison}, {Madsen}, {Miller}, {Stern}, {Walton}, \&
  {Zhang}}]{Fuerst2014}
{F{\"u}rst}, F., {Pottschmidt}, K., {Wilms}, J., {et~al.} 2014, \apj, 780, 133

\bibitem[{{Gnedin} \& {Sunyaev}(1974)}]{Gnedin1974}
{Gnedin}, I.~N. \& {Sunyaev}, R.~A. 1974, \aap, 36, 379

\bibitem[{{Gonthier} {et~al.}(2014){Gonthier}, {Baring}, {Eiles}, {Wadiasingh},
  {Taylor}, \& {Fitch}}]{Gonthier2014}
{Gonthier}, P.~L., {Baring}, M.~G., {Eiles}, M.~T., {et~al.} 2014, \prd, 90,
  043014

\bibitem[{{Greene}(1959)}]{Greene1959}
{Greene}, J. 1959, \apj, 130, 693

\bibitem[{{Harding} \& {Daugherty}(1991)}]{Harding1991}
{Harding}, A.~K. \& {Daugherty}, J.~K. 1991, \apj, 374, 687

\bibitem[{{Herold} {et~al.}(1982){Herold}, {Ruder}, \& {Wunner}}]{Herold1982}
{Herold}, H., {Ruder}, H., \& {Wunner}, G. 1982, \aap, 115, 90

\bibitem[{{Klochkov} {et~al.}(2012){Klochkov}, {Doroshenko}, {Santangelo},
  {Staubert}, {Ferrigno}, {Kretschmar}, {Caballero}, {Wilms}, {Kreykenbohm},
  {Pottschmidt}, {Rothschild}, {Wilson-Hodge}, \&
  {P{\"u}hlhofer}}]{Klochkov2012}
{Klochkov}, D., {Doroshenko}, V., {Santangelo}, A., {et~al.} 2012, \aap, 542,
  L28

\bibitem[{{Kong} {et~al.}(2021){Kong}, {Zhang}, {Ji}, {Reig}, {Doroshenko},
  {Santangelo}, {Staubert}, {Zhang}, {Soria}, {Chang}, {Chen}, {Wang}, {Tao},
  \& {Qu}}]{Kong2021}
{Kong}, L.~D., {Zhang}, S., {Ji}, L., {et~al.} 2021, \apjl, 917, L38

\bibitem[{{Kylafis} {et~al.}(2014){Kylafis}, {Tr{\"u}mper}, \&
  {Ertan}}]{Kylafis2014}
{Kylafis}, N.~D., {Tr{\"u}mper}, J.~E., \& {Ertan}, {\"U}. 2014, \aap, 562, A62

\bibitem[{{Kylafis} {et~al.}(2021){Kylafis}, {Tr{\"u}mper}, \&
  {Loudas}}]{Kylafis2021}
{Kylafis}, N.~D., {Tr{\"u}mper}, J.~E., \& {Loudas}, N.~A. 2021, \aap, 655, A39

\bibitem[{{Landau}(1930)}]{Landau1930}
{Landau}, L. 1930, Zeitschrift fur Physik, 64, 629

\bibitem[{Landau \& Lifshitz(1981)}]{landau1965}
Landau, L. \& Lifshitz, E. 1981, Quantum Mechanics: Non-Relativistic Theory,
  Course of Theoretical Physics (Elsevier Science)

\bibitem[{{Langer} \& {Rappaport}(1982)}]{Langer1982}
{Langer}, S.~H. \& {Rappaport}, S. 1982, \apj, 257, 733

\bibitem[{{Loudas} {et~al.}(2021){Loudas}, {Kylafis}, \&
  {Tr{\"u}mper}}]{Loudas2021}
{Loudas}, N.~A., {Kylafis}, N.~D., \& {Tr{\"u}mper}, J.~E. 2021, \aap, 655, A38

\bibitem[{{Mushtukov} \& {Tsygankov}(2022)}]{Mushtukov2022}
{Mushtukov}, A. \& {Tsygankov}, S. 2022, arXiv e-prints, arXiv:2204.14185

\bibitem[{{Mushtukov} {et~al.}(2016){Mushtukov}, {Nagirner}, \&
  {Poutanen}}]{Mushtukov2016}
{Mushtukov}, A.~A., {Nagirner}, D.~I., \& {Poutanen}, J. 2016, \prd, 93, 105003

\bibitem[{{Mushtukov} {et~al.}(2015{\natexlab{a}}){Mushtukov}, {Suleimanov},
  {Tsygankov}, \& {Poutanen}}]{Mushtukov2015a}
{Mushtukov}, A.~A., {Suleimanov}, V.~F., {Tsygankov}, S.~S., \& {Poutanen}, J.
  2015{\natexlab{a}}, \mnras, 447, 1847

\bibitem[{{Mushtukov} {et~al.}(2015{\natexlab{b}}){Mushtukov}, {Tsygankov},
  {Serber}, {Suleimanov}, \& {Poutanen}}]{Musthukov2015b}
{Mushtukov}, A.~A., {Tsygankov}, S.~S., {Serber}, A.~V., {Suleimanov}, V.~F.,
  \& {Poutanen}, J. 2015{\natexlab{b}}, \mnras, 454, 2714

\bibitem[{{Nishimura}(2014)}]{Nishimura2014}
{Nishimura}, O. 2014, \apj, 781, 30

\bibitem[{{Nobili} {et~al.}(2008){Nobili}, {Turolla}, \& {Zane}}]{Nobili2008}
{Nobili}, L., {Turolla}, R., \& {Zane}, S. 2008, \mnras, 389, 989

\bibitem[{Noebauer \& Sim(2019)}]{Noebauer2019}
Noebauer, U.~M. \& Sim, S.~A. 2019, Living Reviews in Computational
  Astrophysics, 5, 1

\bibitem[{{Postnov} {et~al.}(2015){Postnov}, {Gornostaev}, {Klochkov},
  {Laplace}, {Lukin}, \& {Shakura}}]{Postnov2015}
{Postnov}, K.~A., {Gornostaev}, M.~I., {Klochkov}, D., {et~al.} 2015, \mnras,
  452, 1601

\bibitem[{{Poutanen} {et~al.}(2013){Poutanen}, {Mushtukov}, {Suleimanov},
  {Tsygankov}, {Nagirner}, {Doroshenko}, \& {Lutovinov}}]{Poutanen2013}
{Poutanen}, J., {Mushtukov}, A.~A., {Suleimanov}, V.~F., {et~al.} 2013, \apj,
  777, 115

\bibitem[{{Pozdnyakov} {et~al.}(1983){Pozdnyakov}, {Sobol}, \&
  {Syunyaev}}]{Pozdnyakov1983}
{Pozdnyakov}, L.~A., {Sobol}, I.~M., \& {Syunyaev}, R.~A. 1983, \apspr, 2, 189

\bibitem[{Riffert {et~al.}(1999)Riffert, Klingler, \& Ruder}]{Riffert1999}
Riffert, H., Klingler, M., \& Ruder, H. 1999, Phys. Rev. Lett., 82, 3432

\bibitem[{{Romanova} {et~al.}(2004){Romanova}, {Ustyugova}, {Koldoba}, \&
  {Lovelace}}]{Romanova2004}
{Romanova}, M.~M., {Ustyugova}, G.~V., {Koldoba}, A.~V., \& {Lovelace},
  R.~V.~E. 2004, \apj, 610, 920

\bibitem[{{Rothschild} {et~al.}(2017){Rothschild}, {K{\"u}hnel}, {Pottschmidt},
  {Hemphill}, {Postnov}, {Gornostaev}, {Shakura}, {F{\"u}rst}, {Wilms},
  {Staubert}, \& {Klochkov}}]{Rothschild2017}
{Rothschild}, R.~E., {K{\"u}hnel}, M., {Pottschmidt}, K., {et~al.} 2017,
  \mnras, 466, 2752

\bibitem[{{Schwarm} {et~al.}(2017){Schwarm}, {Sch{\"o}nherr}, {Falkner},
  {Pottschmidt}, {Wolff}, {Becker}, {Sokolova-Lapa}, {Klochkov}, {Ferrigno},
  {F{\"u}rst}, {Hemphill}, {Marcu-Cheatham}, {Dauser}, \&
  {Wilms}}]{Schwarm2017}
{Schwarm}, F.~W., {Sch{\"o}nherr}, G., {Falkner}, S., {et~al.} 2017, \aap, 597,
  A3

\bibitem[{{Shapiro} \& {Salpeter}(1975)}]{Shapiro1975}
{Shapiro}, S.~L. \& {Salpeter}, E.~E. 1975, \apj, 198, 671

\bibitem[{{Sharma} {et~al.}(2022){Sharma}, {Jain}, \& {Dutta}}]{Sharma2022}
{Sharma}, P., {Jain}, C., \& {Dutta}, A. 2022, \mnras, 513, L94

\bibitem[{{Sharma} {et~al.}(2023){Sharma}, {Jain}, \& {Dutta}}]{Sharma2023}
{Sharma}, P., {Jain}, C., \& {Dutta}, A. 2023, \mnras, 522, 5608

\bibitem[{Shui {et~al.}(2024)Shui, Zhang, Wang, Mushtukov, Santangelo, Zhang,
  Kong, Ji, Chen, Doroshenko, Frontera, Chang, Peng, Yin, Qu, Tao, Ge, Li, Ye,
  \& Li}]{Shui2024}
Shui, Q.~C., Zhang, S., Wang, P.~J., {et~al.} 2024, \mnras, stae352

\bibitem[{{Sina}(1996)}]{Sina1996}
{Sina}, R. 1996, PhD thesis, University of Maryland, College Park

\bibitem[{{Staubert} {et~al.}(2007){Staubert}, {Shakura}, {Postnov}, {Wilms},
  {Rothschild}, {Coburn}, {Rodina}, \& {Klochkov}}]{Staubert2007}
{Staubert}, R., {Shakura}, N.~I., {Postnov}, K., {et~al.} 2007, \aap, 465, L25

\bibitem[{{Staubert} {et~al.}(2019){Staubert}, {Tr{\"u}mper}, {Kendziorra},
  {Klochkov}, {Postnov}, {Kretschmar}, {Pottschmidt}, {Haberl}, {Rothschild},
  {Santangelo}, {Wilms}, {Kreykenbohm}, \& {F{\"u}rst}}]{Staubert2019}
{Staubert}, R., {Tr{\"u}mper}, J., {Kendziorra}, E., {et~al.} 2019, \aap, 622,
  A61

\bibitem[{{Truemper} {et~al.}(1978){Truemper}, {Pietsch}, {Reppin}, {Voges},
  {Staubert}, \& {Kendziorra}}]{Truemper1978}
{Truemper}, J., {Pietsch}, W., {Reppin}, C., {et~al.} 1978, \apjl, 219, L105

\bibitem[{{Tr{\"u}mper} {et~al.}(1977){Tr{\"u}mper}, {Pietsch}, {Reppin},
  {Sacco}, {Kendziorra}, \& {Staubert}}]{Truemper1977}
{Tr{\"u}mper}, J., {Pietsch}, W., {Reppin}, C., {et~al.} 1977, Mitteilungen der
  Astronomischen Gesellschaft Hamburg, 42, 120

\bibitem[{{Tsygankov} {et~al.}(2006){Tsygankov}, {Lutovinov}, {Churazov}, \&
  {Sunyaev}}]{Tsygankov2006}
{Tsygankov}, S.~S., {Lutovinov}, A.~A., {Churazov}, E.~M., \& {Sunyaev}, R.~A.
  2006, \mnras, 371, 19

\bibitem[{{Tsygankov} {et~al.}(2010){Tsygankov}, {Lutovinov}, \&
  {Serber}}]{Tsygankov2010}
{Tsygankov}, S.~S., {Lutovinov}, A.~A., \& {Serber}, A.~V. 2010, \mnras, 401,
  1628

\bibitem[{{Vybornov} {et~al.}(2018){Vybornov}, {Doroshenko}, {Staubert}, \&
  {Santangelo}}]{Vybornov2018}
{Vybornov}, V., {Doroshenko}, V., {Staubert}, R., \& {Santangelo}, A. 2018,
  \aap, 610, A88

\bibitem[{{Vybornov} {et~al.}(2017){Vybornov}, {Klochkov}, {Gornostaev},
  {Postnov}, {Sokolova-Lapa}, {Staubert}, {Pottschmidt}, \&
  {Santangelo}}]{Vybornov2017}
{Vybornov}, V., {Klochkov}, D., {Gornostaev}, M., {et~al.} 2017, \aap, 601,
  A126

\bibitem[{{Yang} {et~al.}(2023){Yang}, {Wang}, {Liu}, {Chen}, {Wu}, {Tian}, \&
  {Chen}}]{Yang2023}
{Yang}, W., {Wang}, W., {Liu}, Q., {et~al.} 2023, \mnras, 519, 5402

\bibitem[{{Zhang} {et~al.}(2022){Zhang}, {Blaes}, \& {Jiang}}]{Zhang2022}
{Zhang}, L., {Blaes}, O., \& {Jiang}, Y.-F. 2022, \mnras, 515, 4371

\end{thebibliography}

	\begin{appendix}

        \onecolumn
		\section{Thermally averaged cross-section} \label{App.A}
\begin{figure*}
        \centering
        \includegraphics[width=17cm]{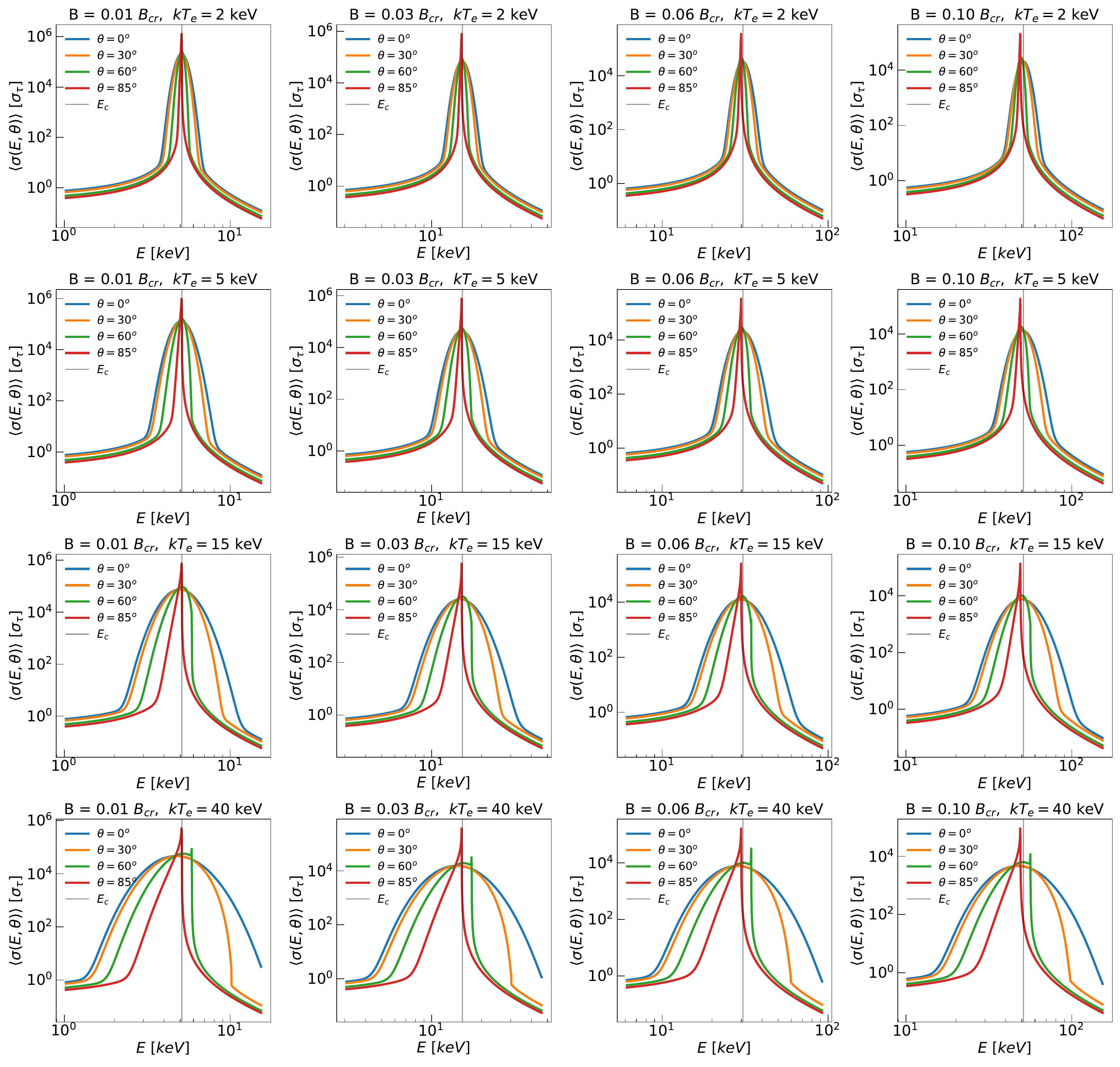}
        \caption{Thermally- and polarization-averaged scattering cross-section as a function of the incident photon energy for various magnetic field strengths, electron temperatures, and incident angles. Each subplot refers to a specific $(B,kT_e)$ set. In each subplot, different line colors correspond to different incident photon angles. Vertical gray line indicates the classical cyclotron energy. }
        \label{fig10}
        \end{figure*}

		This Appendix offers multiple plots of the thermally- and polarization-averaged scattering cross-section used in our Monte Carlo code for the scattering of photons by thermal electrons. To compute the thermally averaged cross-section, we use the polarization-averaged cross-section shown in Eq. \eqref{3.13}. We substitute it into Eq. \eqref{3.5} and average over the electron's velocity along the magnetic field axis, employing a relativistic Maxwellian probability distribution function (see \ref{3.1}). In Fig. \ref{fig10}, we plot the thermally averaged cross-section for a wide range of magnetic field strengths $B$, electron temperatures $kT_e$, and angles of incidence $\theta$. As can be inferred, the width of the cross-section increases remarkably for higher temperatures, which is a natural consequence of the thermal electrons' motion. This implies that magnetic resonant Compton scattering of photons with thermal electrons in the post-shock region can occur for photons that have energy that lies in a larger energy range, thereby leading to a wider absorption feature (thermal broadening). At the same time, though, the peak value decreases since the scattering cross-section is averaged over a normalized probability distribution function. The peak value also decreases for higher magnetic field strengths as the cyclotron line widths $\Gamma_{\pm}$ scale with the magnetic field (see Eqs. \ref{3.11}, \ref{3.12}). For a more detailed analysis of the thermally averaged cross-section, we point the reader to the recent work of \cite{Schwarm2017}.

	\end{appendix}

\end{document}